\documentclass[12pt,a4paper]{article}
\usepackage[utf8]{inputenc}
\usepackage{amsmath,amssymb,amsthm}
\usepackage{xcolor}
\usepackage{tikz}
\usepackage{subcaption}
\usepackage{enumerate}
\usetikzlibrary{decorations.pathmorphing}
\usetikzlibrary{decorations.pathreplacing,angles,quotes}
\usepackage{fontawesome5}
\usepackage[top=1in, left=1in, right=1in, bottom=1in]{geometry}
\usepackage[onehalfspacing]{setspace}
\usepackage{hyperref}
\usepackage{footmisc}
\usepackage{dsfont}

\newcommand{\E}{\mathbb{E}}

\newcommand{\R}{\mathbb{R}}
\newcommand{\Report}{\mathcal{R}}
\newcommand{\N}{\ensuremath{\mathbb{N}}}
\newcommand{\Mna}{M_{(\sigma^\dagger,\rho^\dagger)}}
\newcommand{\Meq}{M_{(\sigma,\rho)}}
\newcommand{\betrag}[1]{\left| #1 \right|}
\newcommand{\m}{\mathfrak{m}}

\newcommand{\one}{\mathds{1}}

\DeclareMathOperator*{\argmax}{arg\,max}
\DeclareMathOperator*{\argmin}{arg\,min}
\DeclareMathOperator{\conv}{conv}

\DeclareMathOperator{\MEU}{MEU}
\DeclareMathOperator{\Bayes}{Bayes}

\newtheorem{proposition}[]{Proposition}
\newtheorem{theorem}[]{Theorem}
\newtheorem{lemma}[]{Lemma}
\newtheorem{corollary}[]{Corollary}

\theoremstyle{definition}\newtheorem{definition}[]{Definition}
\theoremstyle{definition}\newtheorem{assumption}[]{Assumption}
{\theoremstyle{definition}\newtheorem{remark}[]{Remark}}
{\theoremstyle{definition}\newtheorem{example}[]{Example}}

\usepackage[round]{natbib}
\defcitealias{schwartzstein2021using}{S\&S}

\allowdisplaybreaks

\usepackage[normalem]{ulem}

\title{\vspace{-0.5cm}Strategic communication of narratives}
\author{Gerrit Bauch\thanks{Center for Mathematical Economics, Bielefeld University, PO Box 10 01 31, 33501 Bielefeld, Germany. Email: gerrit.bauch@uni-bielefeld.de.} \hskip5pt and Manuel Foerster\thanks{Center for Mathematical Economics, Bielefeld University, PO Box 10 01 31, 33501 Bielefeld, Germany. Email: manuel.foerster@uni-bielefeld.de.
\\\faIcon{creative-commons}\faIcon{creative-commons-by}\faIcon{creative-commons-nc}\faIcon{creative-commons-sa} %This work first circulated as \cite{bauch2024strategiccommunicationnarratives}. 
The authors thank Chiara Aina, Andreas Blume, Yves Breitmoser, Francesco Conti, Arthur Dolgopolov, Tilman Fries, Fabrizio Germano, Lorenz Hartmann, Ole Jann, Dominik Karos, Navin Kartik, Andreas Kleiner, Marco LiCalzi, Julian Matthes, Fynn Närmann, Jurek Preker, Frank Riedel, Ernesto Rivera Mora, Ina Taneva and seminar participants at Bielefeld University, Université Paris Panthéon-Assas, Humboldt University, Paderborn University, the Lisbon Meetings in Game Theory, and at SAET 2025 for fruitful discussions. %
Gerrit Bauch gratefully acknowledges financial support by the German Research Foundation (DFG) [RTG 2865/1 – 492988838]. %
Manuel Foerster gratefully acknowledges financial support from the German Research Foundation (DFG) under grant FO 1272/2-1. %
A sagemath equilibrium solver is available at \href{https://github.com/gbauch/narratives/}{https://github.com/gbauch/narratives/}.}}

\date{\today%\\\bigskip
}

\begin{document}
\maketitle

\begin{abstract}
\noindent%
%127 words
We model the communication of narratives as a cheap-talk game under model uncertainty. The sender has private information about the true data-generating process of public data. The receiver is uncertain about how to interpret the data, but aware of the sender's incentives to strategically provide interpretations (\emph{``narratives''}). We introduce a general class of ambiguity rules resolving the receiver's residual uncertainty. Equilibria are characterized by partitions of narratives into intervals (\emph{``coarse narratives''}). 
In particular, narratives are pooled not according to the similarity of their causal stories but according to the similarity of the optimal receiver actions associated with them.
Furthermore, we establish that public data are difficult to `explain' when some narratives imply similar yet distinct actions. Finally, the persuasive power of the sender is weaker in the sense of state-wise dominance than with a na\"ive receiver.
\end{abstract}

{\footnotesize{\it Keywords:} Narratives, model uncertainty, ambiguity, strategic communication, cheap-talk game

 }

\section{Introduction}
Many important policy decisions are made after communication between an informed adviser and a decision maker whose objectives are not perfectly aligned. Such communication is inherently strategic and often concerns competing interpretations, or \emph{narratives}, about publicly available data. For concreteness, consider a hawkish central bank adviser who briefs a dovish monetary policymaker before an interest-rate decision.\footnote{It is common for central banks to draw on the analysis of external parties, including ``the interpretation of
economic developments by [\ldots] private sector analysts, academic economists, government departments, and industry contacts across the country'', see, e.g., \cite{macklem2002information}.} Both observe the same stream of public releases of macroeconomic indicators. Yet, the relevance of these indicators for assessing the underlying  medium-term inflationary pressure depends on the broader economic circumstances that generated them.

The adviser has access to proprietary information regarding these circumstances and chooses which narrative to communicate, what to disclose or emphasize, and what to leave unresolved, taking into account the policy consequences of her message.\footnote{According to the then-chief of the research department at the Bank of Canada \cite{macklem2002information}, ``[i]nformation on economic activity gathered from industry contacts %across the country 
[\dots] gives the Council insight into [\ldots]
%what business people are seeing and planning and provides insight into
the real-world stories and business decisions that underlie the official statistics.''} In turn, the policymaker interprets that message in light of both the adviser’s informational advantage and her incentives to influence the degree of monetary tightening. The scope for influence in such settings is consequently an equilibrium object: it depends not only on what the adviser knows and prefers, but also on how the decision maker anticipates and responds to the adviser’s strategic framing of the data.

%brief introduction, novelty, teaser
This work develops a novel cheap-talk game under model uncertainty that captures the strategic communication of narratives between a biased sender (she) and an unbiased receiver (he). In contrast to \citet[][henceforth \citetalias{schwartzstein2021using}]{schwartzstein2021using}, we assume that the receiver is aware of the sender's strategic incentives to manipulate him, and we introduce a general class of ambiguity rules to resolve any remaining uncertainty. 
This consequently limits the sender's persuasive power, effectively restricting the set of actions that she can induce.

%model
In our framework, both agents have a common prior belief about the true state of nature $\theta$, interpreted as underlying medium-term inflationary pressure in the motivating example. While the sender also knows the exact underlying process that has generated observable data $h$ from $\theta$, the receiver faces \emph{model uncertainty} (or \emph{ambiguity}) in the sense that he does not know how to interpret the data. For instance, the broader economic circumstances behind an increase in macroeconomic indicators such as consumer price inflation and wage growth may be unclear, i.e., whether the increase reflects persistent domestic inflationary pressure, or rather a temporary supply disruption or an import-cost shock. Any such interpretation is called a \emph{narrative} or a \emph{model} represented by a likelihood function, providing a probabilistic way to make sense of observed data. The finite set $M$ of feasible models contains the true data-generating process.

The biased sender knows the true model and submits a report (or message) to the receiver, aiming at making him interpret the data $h$ in her favor. Any such message provides a reasoning of how to interpret the data, e.g., portraying the public evidence as consistent with high inflationary pressure, formalizing the strategic use of narratives. %
The receiver then interprets this report: under the sender's strategy, it corresponds to the set of models under which she would send it. If this set contains several models, the message conveys a \emph{coarse narrative} that leaves open which of these models generated the data. The receiver then applies an \emph{ambiguity rule} to resolve the remaining model uncertainty and chooses an action. Communication of narratives thus becomes a game in which both agents strategically seek to maximize their own utility. The novel notion of ambiguity rules enables us to treat different preferences in the face of uncertainty in a unified way, including max-min expected utility \citep{gilboa1989maxmin}, the smooth model of decision making \citep{klibanoff2005smooth} and its special case, the classical Bayesian approach. %
In addition, our framework also lifts the maximum likelihood expected utility, the most prominent preference in the recent literature on narratives (\citetalias{schwartzstein2021using}; \citealp{jain2023informing,aina2023tailored}), to an equilibrium setting.

%brief results
Our analysis characterizes equilibria as partitions of models into intervals. Any such interval $\tilde{M}$ corresponds to a coarse narrative, the indicative meaning of which is ``the true model belongs to $\tilde{M}$''. %
Coarse narratives pool models based not on the similarity of their likelihood functions but on the similarity of their associated optimal receiver actions, i.e., the actions they would induce if they were known to be true. Thus, seemingly very different models may be pooled together simply because their policy implications are similar. %
By means of his ambiguity rule, the receiver assigns each interval $\tilde{M}$ a unique optimal action. Among all equilibria in a given setting, there is a maximum number $N$ of induced actions. We provide an algorithm that constructs an equilibrium with $n-1$ distinct actions given one with $n$ distinct actions, thus proving that for any $1 \leq n \leq N$ there is an equilibrium inducing $n$ distinct actions. 

The size of the conflict of interest between the sender and the receiver controls how much information the sender is willing to reveal. We first show that the maximum number $N$ of induced actions is decreasing in the conflict, and then characterize informativeness thresholds that mark when the most finely granulated information revelation is full disclosure or partial disclosure, respectively. Notably, the former, but not the latter, threshold is independent of how the receiver deals with the residual uncertainty, that is, of the ambiguity rule.

We then investigate the sender's ability to \emph{explain} the public data in the sense that she can credibly reveal which underlying true model has generated the data. It follows that public data $h$ is more difficult to explain than $h'$ if and only if the informativeness threshold regarding full disclosure is smaller under $h$ than under $h'$. Roughly speaking, histories that are difficult to explain are those under which some narratives translate into similar yet distinct policy consequences. Thus, difficulty to reveal the true model need not imply poor decisions. %
Moreover, in a running example based on a modified beta-binomial model, reflecting the monetary-policy example, we find that extreme histories are the most difficult to explain. Notably, partially informative communication survives greater conflicts of interest when public observations are purely inflationary than when they are purely non-inflationary under maximum likelihood expected utility, whereas the opposite is true under max-min expected utility.

%Addressing equilibrium selection by the sender, 
Furthermore, we find that the sender prefers more informative equilibria whenever facing a receiver who resolves uncertainty by maximum likelihood expected utility. We give counterexamples that this is not true if the receiver resolves uncertainty by max-min expected utility or a Bayesian approach.

%comparison naive receiver
Finally, we compare the outcome of equilibrium play under the maximum likelihood rule to the benchmark model of \citetalias{schwartzstein2021using}, in which the receiver acts in a ``na\"ive'' fashion: being offered a narrative of how to interpret the observed data, he adopts the proposed model if it fits the data better than a default model. The receiver thus takes the sender's suggestion at face value and ignores her strategic incentives. 
Unsurprisingly, a sender type who can do better than under truthful communication when facing a receiver who engages in equilibrium play will also be able to do so when facing a na\"ive receiver, as incentive constraints are weaker in the latter case. %
However, we show that the sender has more \emph{persuasive power} facing a na\"ive receiver in an even stronger sense, in the spirit of state-wise dominance: she obtains larger expected utility facing a na\"ive receiver under \emph{any} model. %
Thus, the sender generally prefers to deal with a na\"ive receiver. Interestingly, the receiver does not generally prefer to be rational himself.  He sometimes would prefer to be na\"ive, as tight incentive constraints under equilibrium play imply missing out on some chances to mutually increase expected utility.

Revisiting the monetary-policy example, our results imply that a hawkish central bank adviser who communicates with a dovish and strategically sophisticated policymaker will convey only coarse narratives about the macroeconomic indicators. Such coarse narratives may pool seemingly very different narratives because their policy implications are similar. The policymaker is therefore left with residual uncertainty about the data-generating process underlying the observed indicators; we propose resolving this uncertainty using an ambiguity rule. The policymaker’s recognition of the adviser’s hawkish incentives limits the adviser’s ability to induce her preferred policy relative to a benchmark in which the policymaker fails to account for the adviser’s bias.

%%literature overview
\bigskip\noindent%
{\bfseries Related literature. }%
\cite{shiller2017narrative} pointed out the necessity to study narratives from an economic point of view in order to better understand the effects of factual and non-factual information on decisions. The economic literature has thus far conceptualized narratives in different ways.

%%explanation narratives as likelihood functions
Our work ties up with the approach that views narratives as likelihood functions. These establish a probabilistic link between observable data and the parameter of interest. Most closely related to our work is \citetalias{schwartzstein2021using}. The authors let a biased narrator provide narratives to a receiver, aiming at making observable data plausible in her favor. The receiver initially considers a default model, which can in fact be the true data-generating process, but adopts the narrative if it fits the observable data better than his default model. The receiver thus does not consider the strategic incentives of the sender. As a main result of theirs, the persuader finds it easier to manipulate the receiver's beliefs if the default model fits the data poorly. %
\cite{aina2023tailored} considers a similar setting but assumes the persuader commits to a set of narratives before the public data has been realized, while \cite{jain2023informing} introduces a third party who can strategically restrict the narrator's pool of feasible models. In a similar vein, \cite{eliaz2021strategic} allow the sender to strategically provide selective interpretations for her own messages. In a political-economy framework, \cite{bilotta2024coarse} study dynamic electoral competition under coarse voter memory, showing how politicians exploit ambiguity in causal attribution by supplying plausible narratives. %
In a laboratory experiment, \cite{barron2024narrative} find evidence that the likelihood of a narrative is a key determinant for persuasiveness and that individuals use this fact to tailor narratives to their own benefit.

Another way of formalizing narratives is by interpreting a directed acyclic graph as a causal model. %
\cite{eliaz2020model} employ an equilibrium concept of long-run distribution over narrative-policy pairs and show that prevailing models can be misspecified causal models. \cite{eliaz2025false} take this approach to a political-economics framework. %
\cite{foerster2021casting} study communication about externalities in a psychological game, finding that subjects downplay social impact to cast doubt on the effectiveness of action to excuse passivity; see \cite{benabou2020narratives} for a closely related theoretical model.

%cheap talk
Our model combines the literature on narratives as likelihood functions with the one on cheap-talk games. Analogously to the seminal work of \cite{crawford1982strategic}, the biased sender strategically partitions unknown information into intervals.
While they do this for the infinite state space itself, assuming Bayesian agents, we do so for the discrete model space and introduce a general class of ambiguity rules to resolve the receiver's uncertainty, including the Bayesian approach. %
The combination of a cheap-talk game with ambiguity relates our paper to \cite{colo2024communicating}, who assumes the receiver's model uncertainty concerns the prior distribution of the state of the world directly instead of through the interpretation of public data. \citeauthor{colo2024communicating} thus considers cheap-talk communication over a classical interval of beliefs, while we study communication over a discrete set of narratives, making none of his major theorems carry over to our set-up.\footnote{See the discussion below Proposition \ref{Proposition: MLEU equilibria prefer informativeness} on p.\ \pageref{Proposition: MLEU equilibria prefer informativeness:discussion} for details.}

%uncertainty
The importance of uncertainty when it comes to narratives has thus far been acknowledged by researchers on climate change as well as psychologists \citep{pedde2019bridging,constantino2021decision,garcia2010framing}. %
%%formal
We make ambiguity an integral part of our economic model. %
To resolve any remaining uncertainty in this setting, agents need a decision rule to determine their optimal action. While \cite{colo2024communicating} separately studies the pessimistic max-min rule \citep{gilboa1989maxmin} and a Bayesian approach, we introduce ambiguity rules as a unified framework to study a vast class of decision rules under uncertainty. Apart from the max-min rule and the Bayesian approach, it includes the maximum likelihood rule, dominant in the economic literature on narratives, as well as the smooth rule proposed by \cite{klibanoff2005smooth}, which allows for different ambiguity attitudes.

To our knowledge, we are the first to model the communication of narratives as a cheap-talk game under model uncertainty, formalizing narratives as likelihood functions and introducing a general class of ambiguity rules, while the receiver takes the sender's strategic incentives into account.

%%brief overview
The remainder of the article is organized as follows. Section \ref{Section: model and notation} introduces the formal set-up of the game theoretic model, defines ambiguity rules and the equilibrium concept, and provides the running example. Section \ref{Section: General analysis} characterizes equilibria and the informativeness thresholds.
Within the running example, we provide further insights into the informativeness thresholds in Section \ref{Section: conrete analysis Example}. Section \ref{Section: comparison naive} classifies the persuasive power of the sender by comparing the equilibrium model to the one with a na\"ive receiver. Section \ref{Section: Conclusion} concludes and discusses some of our modeling assumptions.

\section{Model and notation}\label{Section: model and notation}

Two agents, a \emph{sender} ($S$ or she) and a \emph{receiver} ($R$ or he), both observe a history of past outcomes (or a public signal) $h\in H$ about the state of nature $\theta\in\Theta=[0,1]$. We assume that $H$ is finite. The common prior over the state is a distribution $F_0$ on $\Theta$ with continuous and strictly positive density $f_0$.
A \emph{model} or \emph{narrative} $m\in M$ parameterizes a likelihood function $\{\pi_m(\cdot|\theta)\}_{\theta\in\Theta}$, where $\pi_m(h|\theta)$ denotes the likelihood of history $h$ given state $\theta$ under model $m$. The \emph{expected fit} of a model $m$ given data $h$ is $\Pr(h \mid m, F_0) := \int_{\Theta} \pi_m(h \mid \theta) \, \mathrm{d} F_0(\theta)$. We assume that $M$ is a finite set and under all $m\in M$ every history $h\in H$ has positive probability given the prior, i.e., $\Pr(h \mid m, F_0)>0$.

The state $\theta$ is drawn according to $F_0$ while there is (model) uncertainty about the \emph{true model} $m^T$. %
The history $h$ is generated according to the true model $m^T$ and the state $\theta$, and observed by both agents. While the sender also learns the true model $m^T$ and updates her prior to the posterior $F_{m^T,h}$ using Bayes' rule, the receiver faces full model uncertainty about $m^T$.\footnote{C.f.\ \cite{muraviev2017kuhn} for a way to model extensive form games under ambiguity. The receiver's initial belief can formally be modeled by a \emph{capacity} $\mu_0$ \citep{choquet1954theory} such that $\mu_0(M')=1$ if $M'=M$ and $\mu_0(M')=0$ if $M'\subsetneq M$, reflecting full uncertainty.\label{footnote: capacity belief}}

After the sender has observed the history $h$ and the true model $m^T$, she submits a cheap-talk report (or message) $r\in \Report$, where $\Report$ is a rich message space.\footnote{A message space in our setting is rich if it has at least $\# M$ elements. Hence, communication is not artificially restricted and allows for a perfect discrimination between models.} Upon observing the report $r$, the receiver can restrict the set of considered models to a \emph{minimal feasible set} $\tilde M$. $\tilde{M}$ is the smallest collection of models the receiver can and does not want to exclude from his considerations, i.e., the indicative meaning of the submitted report is ``$m^T\in\tilde M$''. %
Finally, the receiver takes an action $a\in A=\R$.

\subsection{Payoffs}

The sender's payoff function $u_S(a,\theta,b)$ depends on the receiver's action $a\in A$, the state $\theta\in \Theta$, and the sender's bias $b>0$, a commonly known constant that measures the conflict of interest between sender and receiver. We assume that $u_S(a,\theta,b)$ is twice continuously differentiable and strictly concave in $a$, with a unique maximizer for fixed $\theta$ and $b$. %
Furthermore, we impose \emph{supermodularity}, that is $\frac{\partial^2 u_S(a,\theta,b)}{\partial a\partial b}>0$, a sorting condition common in cheap talk that ensures that the sender would like to induce a higher action with a higher bias. The receiver is considered unbiased and his payoff function is $u_R(a,\theta) := u_S(a,\theta,0)$.

Note that the expected utility function $a \mapsto U_{m,h}(a, b) := \E[u_S(a, \theta, b) \mid m,h]$ after an update on the information revealed through the datum $(m,h)$ is then also strictly concave in $a$ and its unique maximizer, denoted by $a(m,h,b)$, is strictly increasing in $b$. %
For any fixed $h$, we assume \emph{consistent ordering} across biases, meaning that both agents share the ordering of bliss points across models.%
\begin{assumption}[Consistent ordering]
    For any history $h$, models $m,m'$ and bias $b$, we have $a(m',h,0)\leq a(m,h,0)$ if and only if $a(m',h,b)\leq a(m,h,b)$.
\end{assumption}
While sender and receiver may disagree about the appropriate action under any given model, they  agree about which models require a relatively higher action. The sender and receiver thus agree on the order of models defined by
    \begin{equation*}
        m' \leq m \text{ if and only if } a(m',h,0) \leq a(m, h, 0).
    \end{equation*}
We take strict single crossing of the expected utility functions w.r.t.\ models and actions as our second assumption.
\begin{assumption}[Strict single crossing]
    For any history $h$, bias $b$, models $m<m'$, and actions $a<a'$, we have $U_{m',h}(a',b)>U_{m',h}(a,b)$ whenever $U_{m,h}(a',b)\geq U_{m,h}(a,b)$.
\end{assumption}
Single crossing assumptions are common in many cheap-talk, observational learning, and collective choice settings, and closely related to the concepts of interval choice or convex choice, cf.\ \cite{kartik2024single,kartik2024convexchoice}. In particular, strict single crossing implies that best reply sets are \emph{intervals}, where we call $M'\subseteq M$ an {interval} if for all $m,m',m'' \in M$ with $m<m'<m''$ and $m,m'' \in M'$, we also find $m' \in M'$.

Dating back to at least \cite{crawford1982strategic}, the most pervasive utility function in communication games is the \emph{quadratic loss functional}. It fulfills consistent ordering and strict single crossing, irrespective of any assumptions on $M$ and its associated likelihood functions.
\begin{example}\label{Example: quadratic loss}
    Let $u_S(a,\theta,b) = -(\theta+b-a)^2$, which is a strictly concave function in $a$ and supermodular. Under any history $h$ and model $m$, the bliss point is $a(m,h,b)=\E[\theta \mid m,h] + b$. That is, bliss points are ordered by their expected state across biases and thus consistently ordered. %
    Strict single crossing holds because $\E[u_S(a,\theta,b) \mid m,h]$ is strictly concave and depends only on the Euclidean distance of $a$ to $\E[\theta\mid m,h]+b$.
\end{example}

\subsection{Ambiguity rules}
After receiving the sender's message, the receiver still faces uncertainty in form of the minimal feasible set $\tilde{M}$. To resolve the remaining uncertainty, the receiver may adopt any utility function of the following class.

\begin{definition}[Ambiguity rules]\label{def: ambiguity}
    An \emph{ambiguity rule} (for the receiver) is a function $U \colon 2^M \times H \times A \to \R, (\tilde{M}, h, a) \mapsto U_{\tilde{M},h}(a)$, assigning a utility level to each action $a$ when facing the minimal feasible set $\tilde{M}$ under history $h$ with the following properties:
    \begin{enumerate}[(i)]
        \item \label{def: back to Bayes} $U_{\{ m\}, h} (a) = \E[u_R(a,\theta)\mid m,h]$,
        \item \label{def: ambiguity concave and hedging} $U_{\tilde{M}, h}(\cdot)$ has a unique maximizer $a(\tilde{M},h)$ %on $A$ 
        such that for all disjoint $\emptyset \neq M_1,M_2 \subseteq M$ we have $a(M_1 \cup M_2,h) \in \conv(a(M_1,h), a(M_2, h))$.
    \end{enumerate}
\end{definition}
A decision maker takes an ambiguity rule as a measure for the anticipated utility. We require that such a rule is based on Bayesian expected utilities (part \eqref{def: back to Bayes}) and retains a bliss point that hedges against increasing ambiguity by diversification (part \eqref{def: ambiguity concave and hedging}).\footnote{The hedging property is related to weak notions of uncertainty aversion in the ambiguity rules studied below: for MLEU and MEU and given $h, \tilde{M}$, interpret an action as the utility-profile-act $U_{\tilde{M},h}(a)$, which is a concave function and thus fulfills Axiom A.5. in \citet{gilboa1989maxmin}. A smooth rule with concave ambiguity index fulfills \emph{smooth ambiguity aversion} \citep{klibanoff2005smooth}.} %
Our novel notion of ambiguity rules covers a large scope of preferences prominent in the literature on decision theory under uncertainty and allows us to study equilibrium properties in our framework in a unified way.

\begin{example}\label{receiver_pref}
Consider the minimal feasible set $\tilde{M}$ given a history $h$.
\begin{enumerate}[(i)]
  \item\label{receiver_pref:i} \emph{Maximum likelihood expected utility (MLEU)}. Let $\succ_{M,h}$ be any \emph{strict} ordering on $M$ that respects the expected fit, i.e., $m \succ_{M,h} m'$ implies $\Pr(h \mid m, F_0) \geq \Pr(h \mid m',F_0)$. The {maximum-likelihood expected utility (w.r.t.\ $\succ_{M,h}$)} is given by the expected utility
\begin{equation*}
U_{\tilde{M}, h}(a)= \E[u_R(a, \theta) \mid \tilde{m},h] = \int_0^1 u_R(a,\theta)dF_{\tilde m,h}(\theta),
\end{equation*}
where $\tilde{m}$ is the largest element in $\tilde{M}$ w.r.t.\ $\succ_{M,h}$. %
Under the MLEU ambiguity rule, the receiver maximizes his expected utility w.r.t.\ a narrative from the minimal feasible set $\tilde M$ that is most likely to explain the observed data $h$. The strict ordering $\succ_{M,h}$ serves as a tiebreaker.\footnote{Using a tiebreaker is important for two reasons. First, without tiebreaking, the hedging property \eqref{def: ambiguity concave and hedging} of an ambiguity rule might fail. Second, and in contrast to the literature, we cannot resolve ties by applying a ``sender-preferred'' action as the receiver faces ambiguity about the true model and thus about the sender's expected utility. Note also that a choice function $C_h \colon 2^M \to M$, $C_h(\tilde{M}) \in \argmax_{ m \in \tilde{M}} \Pr(h \mid m,F_0)$ with the property that if $M_1 \subseteq M_2$ and $C_h(M_2) \in M_1$, then $C_h(M_1) = C_h(M_2)$ would serve the same purpose as the strict ordering.}
  \item\label{receiver_pref:ii} \emph{Max-min expected utility (MEU)}. The {max-min expected utility} of \cite{gilboa1989maxmin} is given by
\begin{align*}
U_{\tilde{M}, h}(a)=\min_{m \in \tilde M} \E[u_R(a,\theta) \mid m,h]= \min_{m\in \tilde M}\int_0^1 u_R(a,\theta)dF_{m,h}(\theta).
\end{align*}
Under the MEU ambiguity rule, the receiver seeks to maximize his worst-case expected utility given $\tilde M$.

\item \emph{Smooth model}. The {smooth model} of decision making under ambiguity of \cite{klibanoff2005smooth} is given by
\begin{equation*}
    U_{\tilde{M},h}(a) =  \phi^{-1} \left( \sum_{m \in \tilde{M}} \upsilon(m \mid \tilde{M},h) \cdot \phi \left(\E[u_R(a,\theta) \mid m,h ] \right) \right),
\end{equation*}
where we take a strictly increasing and concave ambiguity index $\phi \colon \R \to \R$ and a probability measure $\upsilon$ with full support on $M$. By Bayes' rule, the conditional probability is $\upsilon(m \mid \tilde{M},h) = \frac{\one_{\tilde M}(m)\upsilon(m)\Pr(h \mid m, F_0)}{\sum_{\tilde{m} \in \tilde{M}} \upsilon(\tilde{m})\Pr(h \mid \tilde m, F_0)}$. Note that the smooth model becomes the standard Bayesian expected utility with prior $\upsilon$ if we use a linear ambiguity index $\phi$.
\end{enumerate}
\end{example}

\subsection{Equilibrium concept}\label{subsection:model_eq_concept}

A strategy for the sender is a function $\sigma \colon H \times M \to \Report$ that assigns a report to each history and model. A strategy for the receiver is a function $\rho \colon H \times \Report \to A$ that assigns an action to each history and report. Note that the history $h$ is known to both agents ex ante, i.e., before they make their move. Consequently, one can treat every $h$ as indexing a different game.\footnote{While the restriction to an $h$-path is not a proper subgame as it cuts information sets, equilibria $(\overline{\sigma}, \overline{\rho})$ of the whole game and the considered ``$h$-dependent equilibria'' $(\sigma_h, \rho_h)$ stand in a one-to-one correspondence: fixing any $(\sigma_h,\rho_h)$ for each $h$, we can define $\overline{\sigma}(m,h):=\sigma_h(m)$ and $\overline{\rho}(r,h) := \rho_h(r)$ to obtain an equilibrium. %
Vice versa, given $(\overline{\sigma}, \overline{\rho})$, we can define $\sigma_h(m) := \overline{\sigma}(m,h)$ and $\rho_h(r) := \overline{\rho}(r,h)$ to obtain an $h$-dependent equilibrium.} In the following, we thus omit the dependence of the agents' strategies on $h$ and consider $\sigma \colon M \to \Report$ and $\rho \colon \Report \to A$ for a given history $h$.

Upon observing a report $r \in \sigma(M)$, and given the sender strategy $\sigma$, the receiver can discard all models under which $r$ is not sent. The minimal feasible set is thus $\tilde{M} = \sigma^{-1}(r)$ and the indicative meaning of message $r$ is ``$m^T\in\tilde{M}$''.\footnote{Following footnote \ref{footnote: capacity belief} on p.\ \pageref{footnote: capacity belief}, we can use the \emph{Dempster-Shafer updating rule} \citep{dempster1968generalization,shafer1976mathematical} or the full Bayesian updating rule \citep{fagin1991uncertainty} (if $0/0:=1$) to obtain the updated capacity $\mu_1^r$. The minimal feasible $\tilde{M}$ set can be obtained from $\mu_1^r$ since $\mu_1^r(M') = 1$ if $\tilde{M} \subseteq M'$ and $\mu_1^r(M') = 0$ otherwise.}

We restrict attention to pure strategies and assume that the receiver takes an on-equilibrium-path action if he observes an off-equilibrium message, which implies that we can ignore such deviations.\footnote{The restriction to pure strategies is without loss for the receiver, having a unique utility-maximizing action, while the sender may play mixed strategies in a discrete Bayesian framework \citep[see][]{frug2016note}.}

An \emph{equilibrium} $(\sigma,\rho)$ (under ambiguity rule $U_{\tilde{M}, h}$) then needs to entail mutual best replies:
\begin{enumerate}[(i)]
    \item $\sigma(m^T) \in \argmax_{r} U_{m^T,h}(\rho(r),b)$ for all $m^T\in M$,
    \item $\rho(r) = \argmax_{a} U_{\sigma^{-1}(r),h}(a)$ for all $r \in \sigma(M)$.
\end{enumerate}

A sender who observes the true model $m^T$ sends a message $r$ inducing the action most favorable for her out of $\rho(\Report)$. Being aware of the sender's communication strategy $\sigma$ and observing a message $r$, the receiver first updates the considered set of models to the minimal feasible set $\tilde{M} = \sigma^{-1}(r)$. Then, he chooses the action maximizing his utility w.r.t.\ his ambiguity rule. 

An immediate consequence of the cheap-talk setting is the existence of trivial, so-called \emph{babbling equilibria} in which no information is transmitted: the sender's strategy is constant, i.e., $\sigma(m) = r_0$ for all $m$, and the receiver responds optimally by playing the \emph{pooling action} $\rho(r) = a(M,h)$ for any received message.

\subsection{Running example}\label{ex:uniform_random_quadratic}
We conclude this section by introducing our running example. Consider a hawkish central bank adviser (sender) who communicates with a dovish policymaker (receiver) before an interest-rate decision. Underlying medium-term inflationary pressure is summarized by a parameter $\theta \in [0,1]$. Without further information, this inflationary pressure is believed to be drawn from a uniform distribution $F_0 = \mathcal U(0,1)$. A sequence of public releases of macroeconomic data is available. To simplify, the data is summarized by a vector $h=(h_1,h_2,\ldots,h_K) \in \{0,1\}^K$, indicating public observations that are categorized as inflationary ($1$) or non-inflationary ($0$), depending on whether they are above or below an associated policy target. These observations may represent, e.g., releases on consumer price inflation, indicating whether it is above or below the common policy target of 2\%, or employment, indicating whether it is close to full employment.

In this example, a model is identified with a set of data points attributed to underlying inflationary pressure, i.e., $M = 2^{\{1, \dots, K\}}$. The true model $m^T \in M$ indicates the instances in which underlying inflationary pressure contributed to the observed outcome. At all other times, the result is independent of underlying inflationary pressure, reflecting, e.g., a temporary supply disruption or an import-cost shock, so that $h_1,\dots,h_K$ are independent conditional on $\theta$, with
\begin{align}
\Pr(h_k=1\ |\ \theta)=\left\{\begin{array}{cl} \theta, & \mbox{if }k\in m^T\\ \frac{1}{2}, & \mbox{else} \end{array}\right. \text{for all } k=1,2,\ldots, K. \label{ex:uniform_random_quadratic:tag1}
\end{align}
Note that \eqref{ex:uniform_random_quadratic:tag1} constitutes a modified beta-binomial model. The unbiased policymaker is interested in learning the underlying inflationary pressure, but faces uncertainty about which pieces of the public data, such as a high consumer price inflation, low unemployment, or elevated wage growth, are relevant for this assessment. 

The biased adviser, on the other hand, knows the true data-generating process $m^T$, i.e., which of the observed data is truly connected to underlying inflationary pressure, but not the inflationary pressure itself. For example, she may have access to proprietary information regarding the broader economic circumstances, e.g., firm-level information, wage-setting evidence, or non-public statistical information, that reveals whether observed indicators are indicative of the underlying inflationary pressure. Following the leading example in \cite{crawford1982strategic}, her payoff function is $u_S(a,\theta,b)=-(\theta+b-a)^2$, reflecting her preference for a tighter monetary-policy stance than the policymaker’s. She does so by providing a \emph{narrative} that interprets the observed data as evidence about underlying inflationary pressure while also supporting tighter policy. In our framework, the policymaker takes the adviser’s strategic incentives into account, following equilibrium behavior, and can thus counteract the adviser’s hawkish bias.
 
Before proceeding with the equilibrium analysis, we provide some intuition for equilibria in this setting. Figure \ref{Figure: Examples MLEU equilibria} depicts an equilibrium under the MLEU ambiguity rule (henceforth \emph{MLEU equilibrium}) for $b=\tfrac{1}{25}$, $M = 2^{\{1,2,3\}}$, i.e., $K=3$, and the history $h=(1,0,1)$. %
We associate each model $m\in M$ with the receiver's bliss point $a(m,h,0)=\E[\theta \mid m,h]$ (cf.\ Example \ref{Example: quadratic loss}). As the communication is about narratives, the sender partitions the set of models into intervals, sending the same coarse narrative for all models in a given partition element. %
Under the MLEU ambiguity rule, the receiver picks the model with the highest expected fit from each partition element. Given this action profile and the sender's bias, the sender has no incentives to deviate by sending a different message. For instance, if $m^T=\{1,2\}$ such that the sender's bliss point is $a(m^T,h,b)=a(m^T,h,0)+b=\tfrac{27}{50}$, hence the sender is better off sending $\sigma(m^T)=r_2$ than $r_3$, as
\begin{equation*}
    \left|a(m^T,h,b)-\rho(r_2)\right|=b<\tfrac{3}{50}=\left|a(m^T,h,b)-\rho(r_3)\right|.
\end{equation*}
A central insight of our model is that the sender does not pool narratives 
%based on the similarity between their causal stories, but based on the similarity between 
according to the similarity of their causal stories but according to the similarity of the policy implications associated with them. Indeed, the sender pools the narrative under which the first and the second (resp.\ second and third) indicator is relevant for judging inflationary pressure with the narrative under which none of the indicators is relevant (message $r_2$ in Figure \ref{Figure: Examples MLEU equilibria}) but not with any of the narratives under which exactly one of the two indicators is relevant.
\begin{figure}
    \centering
    \begin{tikzpicture}[scale=18]
        %%sigma scope
		%drawings
		\node at (0.3,0.046) [left,shift={(-0.5,0)}]{$\sigma(m)$:};
        \draw (0.3,0) node[left,shift={(-0.5,0)}]{$\rho(r)$:};
        
		\draw (0.3,0)--(0.8,0);
        \def\blisspoints{1/3,1/2,3/5,2/3,3/4}
        \def\sigmapoints{1/2*1/3+1/2*1/2,1/2*1/2+1/2*3/5, 1/2*3/5+1/2*2/3}
        \def\rhopoints{1/3,1/2,3/5, 3/4}
		\foreach \x in \blisspoints{
		      \draw (\x,0)--(\x,-0.02)node[below,scale=0.75]{$\x$};
        }
        \draw (1/3,-0.08)node[scale=0.75,align=center]{0 \faIcon{check-circle}\\1 \faIcon{times-circle}};
        \draw (1/2,-0.08)node[scale=0.75,align=center]{1 \faIcon{check-circle} \ 0 \faIcon{check-circle}\\1 \faIcon{times-circle} \ 0 \faIcon{times-circle}};
        \draw (3/5,-0.08)node[scale=0.75,align=center]{2 \faIcon{check-circle}\\1 \faIcon{times-circle}};
        \draw (2/3,-0.08)node[scale=0.75,align=center]{1 \faIcon{check-circle}\\0 \faIcon{times-circle}};
        \draw (3/4,-0.08)node[scale=0.75,align=center]{2 \faIcon{check-circle}\\0 \faIcon{times-circle}};

        \draw (0.55,0.1)node[]{$h=(1,0,1)$ $\hat{=}$  2 \faIcon{check-circle} 1 \faIcon{times-circle}};
        
        %boundaries
        %(i)sigma
        \foreach \s in \sigmapoints{
                \draw[thick,dashed] (\s,0.055)--(\s,-0.055);
        }
        %messages
        \draw[decoration={brace,raise=10pt},decorate] (1/3-1/60,0) -- node[above=14pt] {$r_{1}$} (1/3+1/60,0);
        \draw[decoration={brace,raise=10pt},decorate] (1/2-1/60,0) -- node[above=14pt] {$r_{2}$} (1/2+1/60,0);
        \draw[decoration={brace,raise=10pt},decorate] (3/5-1/60,0) -- node[above=14pt] {$r_{3}$} (3/5+1/60,0);
        \draw[decoration={brace,raise=10pt},decorate] (2/3-1/60,0) -- node[above=14pt] {$r_{4}$} (3/4+1/60,0);

        %rho
        \foreach \a in \rhopoints{
            \filldraw (\a,0) circle (0.15pt);
        }

		\draw[->] (0.825,-0.03)node[right,align=left,scale=0.85]{receiver's bliss\\ points $a(m,h,0)$} -- (0.8,-0.03);
        \draw[->] (0.825,-0.08)node[right,align=left,scale=0.85]{models} -- (0.8,-0.08);
        %expected fit
        \draw (1/3,-0.12)node[scale=0.85,align=center]{$\frac{1}{8}$};
        %\draw (1/2,-0.12)node[scale=0.85,align=center]{$\frac{1}{12}$ \ \ $\frac{1}{8}$};
        \draw (1/2,-0.12)node[scale=0.85,align=center]{$\frac{1}{12}$ \ \ \ \  $\frac{1}{8}$};
        \draw (3/5,-0.12)node[scale=0.85,align=center]{$\frac{1}{12}$};
        \draw (2/3,-0.12)node[scale=0.85,align=center]{$\frac{1}{8}$};
        \draw (3/4,-0.12)node[scale=0.85,align=center]{$\frac{1}{6}$};
        \draw[->] (0.825,-0.12)node[right,align=left,scale=0.85]{expected fit} -- (0.8,-0.12);        
        
    \end{tikzpicture}
    \caption{MLEU equilibrium in the setting of Section \ref{ex:uniform_random_quadratic}, where $b=\tfrac{1}{25}$, $M = 2^{\{1,2,3\}}$, and $h= (1,0,1)$ entails two successes (\faIcon{check-circle}) and one failure (\faIcon{times-circle}). Reports span consecutive bliss points, each of which corresponds to at least one model. Each model or narrative corresponds to a claimed number of relevant successes and failures and has an associated expected fit. Note that some models are statistically indistinguishable and are not depicted separately. For instance, $m=\{1,2\}$ and $m' = \{2,3\}$ both indicate one relevant success and failure. The optimal receiver responses for each report are indicated by the solid dots.}
    \label{Figure: Examples MLEU equilibria}
\end{figure}

\section{Equilibrium analysis}\label{Section: General analysis}

In this section, we characterize equilibria in the general framework and derive informativeness thresholds on the conflict of interest. In a first step, we state necessary conditions for establishing equilibria. In equilibrium, messages span intervals of models, i.e., messages correspond to sets of narratives the respective bliss points of which are consecutive elements w.r.t.\ the set of all bliss points. This helps reducing the complexity of finding equilibria. Second, our main theorem provides a full characterization of the set of reduced equilibria. If an equilibrium induces $n+1$ different actions, there also is an equilibrium which induces $n$ different actions. Third, we derive thresholds on the conflict of interest that mark when the most finely granulated information revelation is full disclosure or partial disclosure, respectively.

\subsection{Necessary conditions}

Our first result describes important structural properties of the equilibria that help simplify the search for equilibria.

\begin{lemma}[Reduction Lemma]\label{Lemma: reduction lemma}
    Let $(\sigma', \rho')$ be an equilibrium.
    \begin{enumerate}[(i)]
        \item \label{item: reduction lemma rho injective}There is a corresponding \emph{reduced equilibrium} $(\sigma,\rho)$ with $\rho \circ \sigma \equiv \rho' \circ \sigma'$ and whenever $\rho(\sigma(m)) = \rho(\sigma(m'))$ then $\sigma(m) = \sigma(m')$.
        \item \label{item: reduction lemma ordering} We have $\rho'(\sigma'(m')) \leq \rho'(\sigma'(m))$ whenever $m' < m$.
	\end{enumerate}
\end{lemma}
All proofs are relegated to the appendix.

Statement \eqref{item: reduction lemma rho injective} lets us restrict attention to reduced equilibria in which messages and induced actions are in a one-to-one correspondence:\footnote{In that sense, messages can be identified with action recommendations, a property sometimes referred to as \emph{straightforwardness}.} If an equilibrium has two different messages that induce the same action, one does not lose the equilibrium property by sending the same message for all corresponding narratives. %
Statement \eqref{item: reduction lemma ordering} says that equilibrium actions respect the ordering of bliss points. This property is due to strict single crossing. Consequently, a message partitions narratives into intervals w.r.t.\ the ordering of the bliss points. Notably, narratives are partitioned w.r.t.\ the similarity not of their likelihood functions but of their associated receiver bliss points.

Furthermore, it follows from statement \eqref{item: reduction lemma ordering} that the number of distinct receiver bliss points is an upper bound for the number of actions induced in equilibrium.
\begin{corollary}\label{Corollary: upper bound N stage}
An equilibrium $(\sigma,\rho)$ induces at most $\#\{a(m,h,0) \mid m\in M\}$ distinct actions.
\end{corollary}

\subsection{Equilibria}
In the following, we study the existence of equilibria with a certain number of induced actions and characterize the set of equilibria. We categorize equilibria according to the number of induced actions.

\begin{definition}[$n$-step equilibrium]
    An equilibrium $(\sigma, \rho)$ is called \emph{$n$-step equilibrium} if it induces $n$ distinct actions, i.e., $\# (\rho \circ \sigma)(M) = n$.
\end{definition}
Note first that by Corollary \ref{Corollary: upper bound N stage} any equilibrium induces at most $\#\{a(m,h,0) \mid m\in M\}$ distinct actions. Second, a reduced $1$-step equilibrium is a trivial babbling equilibrium, which always exists. Third, by the Reduction Lemma \ref{Lemma: reduction lemma}, the reduced equilibrium corresponding to an $n$-step equilibrium has $n$ distinct messages. We henceforth characterize equilibria by the corresponding reduced equilibria.

A classical result from the theory of cheap-talk games states that if there is an equilibrium which induces $N$ distinct actions, then there is an equilibrium inducing $n$ distinct actions for all $1 \leq n \leq N$, cf.\ \cite{crawford1982strategic}. In their setting, messages correspond to intervals which partition the continuous state space. Surprisingly, this result also holds true in our more complex setting despite that messages correspond to intervals which partition the \emph{finite} model space. Together with the Reduction Lemma \ref{Lemma: reduction lemma}, this characterizes the set of equilibria: an equilibrium is a partition of the set $M$ into up to $N$ intervals together with their actions induced by the considered ambiguity rule. %
Specifically, the partition pools models not according to the similarity of their likelihood functions but according to the similarity of their associated receiver bliss points. We say that a partition of $M$ is consecutive if it is consecutive w.r.t.\ the corresponding bliss points.

\begin{theorem}\label{Theorem: N step equilibrium}
There is a natural number $N=N(h,b)$ such that there exists an $n$-step equilibrium if and only if $1 \leq n \leq N$. For any $n$-step equilibrium, the corresponding reduced equilibrium $(\sigma,\rho)$ is characterized by a consecutive partition $\{M_1,M_2,\ldots,M_n\}$ of $M$ and distinct $r_1,r_2,\ldots,r_n\in \Report$ such that for all $i=1,2,\ldots,n$
\begin{enumerate}[(i)]
  \item $r_i=\sigma(m)$ if $m\in M_i$,
  \item\label{Theorem: N step equilibrium:ii} $\rho$ satisfies $\rho(r_i)=\argmax\limits_{a} U_{M_i,h}(a)$, and
  \item $r_i \in \argmax_{r} U_{m^T,h}(\rho(r),b)$ for all $m^T\in M_i$.
\end{enumerate}
\end{theorem}

By the Reduction Lemma \ref{Lemma: reduction lemma}, the set of narratives inducing an action is an interval in equilibrium. Furthermore, the actions induced in a reduced $n$-step equilibrium (Theorem \ref{Theorem: N step equilibrium} \eqref{Theorem: N step equilibrium:ii}) are pairwise distinct. We can hence index these intervals $M_i$ such that narratives from an interval with a higher index induce a higher action, i.e., $\rho(\sigma(m)) <  \rho(\sigma(m'))$ whenever $m \in M_i$ and $m' \in M_{j}$ for $i<j$. Thus, reduced equilibria are characterized by a consecutive partition of $M$. Note that it is straightforward to reconstruct non-reduced equilibria from a reduced one:
\begin{remark}
Consider a reduced equilibrium $(\sigma,\rho)$ characterized by the consecutive partition $\{M_1,M_2,\ldots,M_n\}$. 
Any $n$-step equilibrium $(\sigma',\rho')$ of which $(\sigma,\rho)$ is the corresponding reduced equilibrium is characterized by a refined partition $\{M_1',M_2',\ldots,M_{n'}'\}$ of $\{M_1,M_2,\ldots,M_{n}\}$ and distinct $r_1',r_2',\ldots,r_{n'}'\in \Report$, $n'\geq n$, such that $r_i'=\sigma'(m)$ if $m\in M_i'$ for all $i=1,2,\ldots,n'$ and $\rho \circ \sigma \equiv \rho' \circ \sigma'$. %If $n'=n$, then $M_i=M_i'$ for all $i=1,\ldots,n$.
\end{remark}
The proof of the theorem then gives at hand an explicit algorithm to arrive at an $n$-step equilibrium given an $(n+1)$-step equilibrium that we explain in the following. 

\paragraph{Algorithm.} Let $(\sigma,\rho)$ be an $(n+1)$-step equilibrium characterized by the consecutive partition $\{M_1,M_2,\ldots,M_{n+1}\}$.

\begin{enumerate}
    \item Define the sender strategy 
    \begin{align*}
     \sigma^1(m) := \begin{cases}
        r_n &, \mbox{ if } m \in M_n^1\\
        r_i &, \mbox{ if } m \in M_i^1, i \neq n,n+1
    \end{cases},
    \end{align*}
    where $M_i^1=M_i$ for $i=1,2,\ldots,n-1$ and $M_n^1=M_n \cup M_{n+1}$.

    \item\label{item: algorithm terminination} If $\sigma^j$ together with the respective best reply $\rho^j$ of the receiver is an $n$-step equilibrium, we are done. Otherwise go to Step 3.
    
    \item\label{item: algorihm deviation} There exists $i\in \{2,\ldots,n\}$ such that the sender has a profitable deviation for a model $m' \in M_i^j$ from $\sigma^j$ by sending the report $r_{i-1}$, where $m'$ is minimal in $M_i^j$, i.e., with smallest receiver bliss point $a(m',h,0)$. Denote by $\m'$ the set of all minimal narratives in $M_i^j$ with this profitable deviation. %
    Let 
        \begin{itemize}
          \item $\sigma^{j+1}$ be such that $\sigma^{j+1}(m) = r_{i-1}$ if $m \in \m'$ and $\sigma^{j+1}(m) = \sigma^{j}(m)$ else,
          \item $M^{j+1}_{i-1} = M^j_{i-1} \cup \m'$, $M^{j+1}_i = M^j_i \setminus \m'$, and $M^{j+1}_k=M^j_k$ for $k\neq i-1,i$. 
        \end{itemize}
         Go back to Step 2.\qed
\end{enumerate}

\noindent%
In the first step of the algorithm, we merge the rightmost messages to obtain $\sigma^1$ which uses $n$ messages. If $\sigma^1$ together with its best reply $\rho^1$ does not form an equilibrium, we iteratively modify the sender's strategy: There is a profitable deviation for the sender by sending the lower report $r_{i-1}$ for a minimal model $m' \in M^j_i$ for some $i$ (in the first step, this is because we have merged the rightmost messages, i.e., $i=n$). Let $\m' \subseteq M^j_i$ denote the set of all such minimal models and modify $\sigma^j$ by assigning these models to report $r_{i-1}$ instead of $r_i$. We check whether the resulting sender strategy $\sigma^{j+1}$ together with the receiver's best reply constitutes an equilibrium. If not, continue as before and note that only deviations are possible that send lower messages. If this procedure continues to produce no equilibrium, the $M_i^j$ will eventually each only contain models with the same bliss point for $2 \leq i \leq n$. At this point, no further profitable deviation of the sender is possible, proving that the algorithm terminates. The key idea is illustrated in Figure \ref{Figure: theorem proof explanation}.

\begin{figure}[htb]
    \centering
    \begin{tikzpicture}[scale=15]

        \begin{scope}
        %%sigma scope
		%drawings
		\draw (0.3,0)--(0.8,0)node[right,shift={(0.3,0)}]{equilibrium};

        %sigma,rho
        \node at (0.3,0.046) [left,shift={(-0.5,0)}]{$\sigma$:};
        \draw (0.3,0) node[left,shift={(-0.5,0)}]{$\rho$:};

        \def\blisspoints{1/3,1/2,3/5,2/3,3/4}
        \def\sigmapoints{1/2*1/3+1/2*1/2,1/2*2/3+1/2*3/4}
        \def\rhopoints{1/3,2/3,3/4}
		\foreach \x in \blisspoints{
		      \draw (\x,0)--(\x,-0.02)node[below,scale=0.75]{$\x$};
        }
        %boundaries
        %(i)sigma
        \foreach \s in \sigmapoints{
                \draw[thick,dashed] (\s,0.055)--(\s,-0.055);
        }

        %rho
        \foreach \a in \rhopoints{
            \filldraw (\a,0) circle (0.15pt);
        }

        %messages
        \draw[decoration={brace,raise=10pt},decorate] (1/3-1/60,0) -- node[above=14pt] {$r_{1}$} (1/3+1/60,0);
        \draw[decoration={brace,raise=10pt},decorate] (1/2-1/60,0) -- node[above=14pt] {$r_{2}$} (2/3+1/60,0);
        \draw[decoration={brace,raise=10pt},decorate] (3/4-1/60,0) -- node[above=14pt] {$r_{3}$} (3/4+1/60,0);
        \end{scope}

        \begin{scope}[shift={(0,-0.15)}]
        %%sigma_a scope
		%drawings
		\draw (0.3,0)--(0.8,0)node[right,shift={(0.3,0)}]{no equilibrium};

        %sigma,rho
        \node at (0.3,0.046) [left,shift={(-0.5,0)}]{$\rightsquigarrow \sigma^1$:};
        \draw (0.3,0) node[left,shift={(-0.5,0)}]{$\rho^1$:};
  
        \def\blisspoints{1/3,1/2,3/5,2/3,3/4}
        \def\sigmapoints{1/2*1/3+1/2*1/2}
        \def\rhopoints{1/3,3/4}
		\foreach \x in \blisspoints{
		      \draw (\x,0)--(\x,-0.02)node[below,scale=0.75]{$\x$};
        }
        %boundaries
        %(i)sigma
        \foreach \s in \sigmapoints{
                \draw[thick,dashed] (\s,0.055)--(\s,-0.055);
        }
        
        %rho
        \foreach \a in \rhopoints{
            \filldraw (\a,0) circle (0.15pt);
        }

        \draw[->] (5/12,0.038)--(5/12+0.05,0.038);

        %messages
        \draw[decoration={brace,raise=10pt},decorate] (1/3-1/60,0) -- node[above=14pt] {$r_{1}$} (1/3+1/60,0);
        \draw[decoration={brace,raise=10pt},decorate] (1/2-1/60,0) -- node[above=14pt] {$r_{2}$} (3/4+1/60,0);
        \end{scope}

        \begin{scope}[shift={(0,-0.3)}]
        %%sigma_a scope
		%drawings
		\draw (0.3,0)--(0.8,0)node[right,shift={(0.3,0)}]{equilibrium};

        %sigma, rho
        \node at (0.3,0.046) [left,shift={(-0.5,0)}]{$\rightsquigarrow\sigma^2$:};
        \draw (0.3,0) node[left,shift={(-0.5,0)}]{$\rho^2$:};        
  
        \def\blisspoints{1/3,1/2,3/5,2/3,3/4}
        \def\sigmapoints{1/2*3/5+1/2*1/2}
        \def\rhopoints{1/3,3/4}
		\foreach \x in \blisspoints{
		      \draw (\x,0)--(\x,-0.02)node[below,scale=0.75]{$\x$};
        }
        %boundaries
        %(i)sigma
        \foreach \s in \sigmapoints{
                \draw[thick,dashed] (\s,0.055)--(\s,-0.055);
        }
        
        %rho
        \foreach \a in \rhopoints{
            \filldraw (\a,0) circle (0.15pt);
        }

        %messages
        \draw[decoration={brace,raise=10pt},decorate] (1/3-1/60,0) -- node[above=14pt] {$r_{1}$} (1/2+1/60,0);
        \draw[decoration={brace,raise=10pt},decorate] (3/5-1/60,0) -- node[above=14pt] {$r_{2}$} (3/4+1/60,0);
        \end{scope}
    \end{tikzpicture}
    \caption{Sketch of the algorithm driving the proof of Theorem \ref{Theorem: N step equilibrium}. Consider $b= \frac{1}{30}$, $M = 2^{\{ 1,2,3 \}}$, $h=(1,0,1)$, and the MLEU ambiguity rule with any tiebreaker that favors the narrative with no success and one failure over the one with one success and no failure, and the latter over the one with no success and no failure. The dots mark the respective best responses of the receiver. Starting from the $3$-step equilibrium $\sigma$, merge the two rightmost messages to obtain $\sigma^1$. A profitable deviation must involve inducing a lower action, shifting the boundary to the right. By doing so, equilibrium $(\sigma^2,\rho^2)$ is reached.}
    \label{Figure: theorem proof explanation}
\end{figure}

The algorithm also yields an intuitive comparative statics result on the conflict of interest. The smaller the conflict of interest, the more the sender is willing to discriminate between different models in terms of the step size in equilibrium.

\begin{corollary}\label{Corollary:comp_stat}
$N(h,b)\leq N(h,b')$ if $b'<b$.
\end{corollary}

\subsection{Informativeness}\label{Section: Informativeness}

We introduce a partial order, grounded in \citet{blackwell1953equivalent}’s seminal notion of informativeness, to rank equilibria by the granularity of the sender-induced action partition.

\begin{definition}[Informativeness]\label{def:informariveness}
\leavevmode
\begin{enumerate}[(i)]
   \item Equilibrium $(\sigma,\rho)$ is \emph{weakly more informative} than  equilibrium $(\sigma',\rho')$ (under $h$) if $\left(\rho \circ \sigma\right)(m)=\left(\rho \circ \sigma\right)(m')$ implies $\left(\rho' \circ \sigma'\right)(m)=\left(\rho' \circ \sigma'\right)(m')$ for all $m,m'\in M$. 
\item An equilibrium $(\sigma,\rho)$ is \emph{most informative} (under $h$) if for any equilibrium $(\sigma',\rho')$ that is weakly more informative than $(\sigma,\rho)$, it holds that $(\sigma,\rho)$ is weakly more informative than $(\sigma',\rho')$.
\item An equilibrium $(\sigma,\rho)$ is \emph{fully informative} (under $h$) if $(\rho \circ \sigma)(m) = a(m,h,0)$ for all $m\in M$. It is \emph{uninformative} if $(\rho \circ \sigma)(m) = a(M,h)$ for all $m\in M$.
    \end{enumerate}

\end{definition}

\noindent%
Some remarks seem in order. First, Definition \ref{def:informariveness} defines a partial order on equilibria. In particular, note that a more informative equilibrium has a weakly larger step size (cf.\ Theorem \ref{Theorem: N step equilibrium}, Corollary \ref{Corollary:comp_stat}). Moreover, a fully informative equilibrium identifies the receiver's bliss points, while an uninformative equilibrium induces the pooling action $a(M,h)$ depending on the ambiguity rule.
Second, Definition \ref{def:informariveness} coincides with Blackwell's informativeness when passing to reduced equilibria:

\begin{remark}\label{Remark: more informative from equilibrium to Blackwell}
Suppose that equilibrium $(\sigma',\rho')$ is weakly more informative than equilibrium $(\tilde \sigma',\tilde \rho')$ and consider the corresponding reduced equilibria $(\sigma,\rho)$ resp.\ $(\tilde \sigma,\tilde \rho)$. Then $\sigma$ is \emph{weakly more Blackwell informative} than $\tilde\sigma$, i.e., $\sigma(m)=\sigma(m')$ implies $\tilde\sigma(m)=\tilde\sigma(m')$ for all $m,m'\in M$. 
\end{remark}

Third, a fully informative equilibrium is most informative. Fourth, there might be several most informative equilibria, neither of which is weakly more informative than the other. %
The last point is illustrated in Figure \ref{Figure: another three step equilibrium} by two reduced $4$-step equilibria.

\begin{figure}
    \centering
    \begin{tikzpicture}[scale=18]
        %%sigma scope
		%drawings
        
        \begin{scope}

        %sigma,rho
		\node at (0.3,0.046) [left,shift={(-0.5,0)}]{$\sigma_1$:};
        \draw (0.3,0) node[left,shift={(-0.5,0)}]{$\rho_1$:};
        
        \draw (0.3,0)--(0.8,0);
        \def\blisspoints{1/3,1/2,3/5,2/3,3/4}
        \def\sigmapoints{1/2*1/3+1/2*1/2,1/2*1/2+1/2*3/5, 1/2*3/5+1/2*2/3}
        \def\rhopoints{1/3,1/2,3/5, 3/4}
		\foreach \x in \blisspoints{
		      \draw (\x,0)--(\x,-0.02)node[below,scale=0.75]{$\x$};
        }
        \draw (1/3,-0.08)node[scale=0.75,align=center]{0 \faIcon{check-circle}\\1 \faIcon{times-circle}};
        \draw (1/2,-0.08)node[scale=0.75,align=center]{1 \faIcon{check-circle} \ 0 \faIcon{check-circle}\\1 \faIcon{times-circle} \ 0 \faIcon{times-circle}};
        \draw (3/5,-0.08)node[scale=0.75,align=center]{2 \faIcon{check-circle}\\1 \faIcon{times-circle}};
        \draw (2/3,-0.08)node[scale=0.75,align=center]{1 \faIcon{check-circle}\\0 \faIcon{times-circle}};
        \draw (3/4,-0.08)node[scale=0.75,align=center]{2 \faIcon{check-circle}\\0 \faIcon{times-circle}};

        \draw (0.55,0.1)node[]{$h=(1,0,1)$ $\hat{=}$  2 \faIcon{check-circle} 1 \faIcon{times-circle}};
        
        %boundaries
        %(i)sigma
        \foreach \s in \sigmapoints{
                \draw[thick,dashed] (\s,0.055)--(\s,-0.055);
        }
        %messages
        \draw[decoration={brace,raise=10pt},decorate] (1/3-1/60,0) -- node[above=14pt] {$r_{1}$} (1/3+1/60,0);
        \draw[decoration={brace,raise=10pt},decorate] (1/2-1/60,0) -- node[above=14pt] {$r_{2}$} (1/2+1/60,0);
        \draw[decoration={brace,raise=10pt},decorate] (3/5-1/60,0) -- node[above=14pt] {$r_{3}$} (3/5+1/60,0);
        \draw[decoration={brace,raise=10pt},decorate] (2/3-1/60,0) -- node[above=14pt] {$r_{4}$} (3/4+1/60,0);

        %rho
        \foreach \a in \rhopoints{
            \filldraw (\a,0) circle (0.15pt);
        }
        \end{scope}

        %%%second scope
        \begin{scope}[shift={(0,-0.2)}]

        %sigma,rho
		\node at (0.3,0.046) [left,shift={(-0.5,0)}]{$\sigma_2$:};
        \draw (0.3,0) node[left,shift={(-0.5,0)}]{$\rho_2$:};
        
        \draw (0.3,0)--(0.8,0);
        \def\blisspoints{1/3,1/2,3/5,2/3,3/4}
        \def\sigmapoints{1/2*1/3+1/2*1/2,1/2*1/2+1/2*3/5,1/2*2/3+1/2*3/4}
        \def\rhopoints{1/3,1/2,2/3,3/4}
		\foreach \x in \blisspoints{
		      \draw (\x,0)--(\x,-0.02)node[below,scale=0.75]{$\x$};
        }
        \draw (1/3,-0.08)node[scale=0.75,align=center]{0 \faIcon{check-circle}\\1 \faIcon{times-circle}};
        \draw (1/2,-0.08)node[scale=0.75,align=center]{1 \faIcon{check-circle} \ 0 \faIcon{check-circle}\\1 \faIcon{times-circle} \ 0 \faIcon{times-circle}};
        \draw (3/5,-0.08)node[scale=0.75,align=center]{2 \faIcon{check-circle}\\1 \faIcon{times-circle}};
        \draw (2/3,-0.08)node[scale=0.75,align=center]{1 \faIcon{check-circle}\\0 \faIcon{times-circle}};
        \draw (3/4,-0.08)node[scale=0.75,align=center]{2 \faIcon{check-circle}\\0 \faIcon{times-circle}};

        %boundaries
        %(i)sigma
        \foreach \s in \sigmapoints{
                \draw[thick,dashed] (\s,0.055)--(\s,-0.055);
        }
        %messages
        \draw[decoration={brace,raise=10pt},decorate] (1/3-1/60,0) -- node[above=14pt] {$r'_{1}$} (1/3+1/60,0);
        \draw[decoration={brace,raise=10pt},decorate] (1/2-1/60,0) -- node[above=14pt] {$r'_{2}$} (1/2+1/60,0);
        \draw[decoration={brace,raise=10pt},decorate] (3/5-1/60,0) -- node[above=14pt] {$r'_{3}$} (2/3+1/60,0);
        \draw[decoration={brace,raise=10pt},decorate] (3/4-1/60,0) -- node[above=14pt] {$r'_{4}$} (3/4+1/60,0);

        %rho
        \foreach \a in \rhopoints{
            \filldraw (\a,0) circle (0.15pt);
        }
        \end{scope}
    \end{tikzpicture}
    \caption{Two most informative reduced MLEU equilibria that induce different model-action profiles in the setting of Section \ref{ex:uniform_random_quadratic} with $b=\tfrac{1}{25}$, $M = 2^{\{1,2,3\}}$, and $h= (1,0,1)$.}
    \label{Figure: another three step equilibrium}
\end{figure}

In the following, we characterize thresholds on the conflict of interest that mark when fully or partially informative communication is possible in equilibrium. By the Reduction Lemma \ref{Lemma: reduction lemma}, a $2$-step equilibrium consists of a partition of the set $M$ into two intervals. Recall that for a subset $\tilde{M}\subseteq M$ the unique optimal receiver action given his ambiguity rule is denoted by $a(\tilde{M},h)$. Define for each \emph{non-maximal} $m^T$, i.e., $a(m^T,h,0) < \max_{m \in M}a(m,h,0)$,
\begin{equation*}
V(m^T,h,b) := U_{m^T,h}(a(\{ m \mid m > m^T \} ,h),b)-U_{m^T,h}(a(\{ m \mid m \leq m^T \}, h), b).
\end{equation*}
There is no $2$-step equilibrium if $V(m^T,h,b)>0$ for all non-maximal $m^T$.\footnote{Technically, we would need to allow different models $m,m'$ with $a(m,h,0) = a(m',h,0)$ to end up in either of the partition elements, which could be achieved by considering all strict refinements of $\leq$ on $M$. This formality is not restrictive in the subsequent analysis.} As a consequence of Theorem \ref{Theorem: N step equilibrium}, the only equilibrium in this case is the babbling equilibrium.

\begin{definition}[Large conflicts of interest]
    We say that $U_{m,h}(a,b)$ \emph{allows for large conflicts of interest} if for any $h$ there exists a $\hat{b}(h) \geq 0$ such that $V(m^T,h,b)>0$ for all $b > \hat{b}(h)$ and all non-maximal $m^T \in M$.
\end{definition}

\noindent%
Under large conflicts of interest, a sufficiently biased sender cannot induce two different actions without having an incentive to deviate: there will always be a true model under which she seeks to make the receiver pick a higher action than is designed. To illustrate this point, imagine that the bias, measuring the conflict of interest, is high enough to want her make the receiver always pick the highest possible action. Then there cannot be a $2$-step equilibrium. Note that under a quadratic loss the bliss points of the sender are translations of $a(m^T,h,0)$ by exactly $b$, thus allowing for large conflicts of interest.

Analogously to standard cheap-talk models in which the sender is imperfectly informed \citep[e.g.,][]{argenziano2016strategic,foerster2023strategic}, we obtain the following \emph{informativeness thresholds} on the 
conflict of interest:
\begin{proposition}\label{pro:fully_inf}
For any history $h$, there exists
\begin{enumerate}[(i)]
  \item\label{pro:fully_inf:i} $\underline b(h)>0$ such that a most informative equilibrium under history $h$ involves a fully informative sender strategy if and only if $b\leq \underline b(h)$, and 
      
  \item\label{pro:fully_inf:ii} $\overline{b}(h)\geq\underline b(h)$, such that any most informative equilibrium under history $h$ is uninformative if and only if $b> \overline{b}(h)$, provided $U_{m,h}(a,b)$ allows for large conflicts of interest.
\end{enumerate}
Furthermore, $\underline b(h)$ is independent of the ambiguity rule.
\end{proposition}

Intuitively, the first informativeness threshold is independent of the ambiguity rule because there is no ambiguity in fully informative communication. The second threshold, however, is not because the optimal receiver actions depend on the ambiguity rule in partially informative communication (see also Figure \ref{fig:running_overline b}).

To illustrate the thresholds, we revisit the running example introduced in Section \ref{ex:uniform_random_quadratic}. In particular, it may indeed be the case that the two informativeness thresholds coincide under the MLEU ambiguity rule:

\begin{example}\label{ex2}
Consider the setting of Section \ref{ex:uniform_random_quadratic}, where $F_0 = \mathcal U(0,1)$, $u_S(a,\theta,b)=-(\theta+b-a)^2$, and $h$ is generated according to \eqref{ex:uniform_random_quadratic:tag1}, and the MLEU ambiguity rule.
\begin{enumerate}[(i)]
  \item\label{ex2:i} If $h=(1,0,1)$, then $\underline b(h)=\frac{1}{30}<\frac{5}{24}=\overline{b}(h)$.
  \item\label{ex2:ii} If $h=(0,0,0)$, then $\underline b(h)=\overline{b}(h)=\frac{1}{40}$ if the receiver breaks the tie in expected fit between the narrative indicating one failure and the one indicating no information in favor of the first one.\footnote{If, instead of $M = 2^{\{1,2,3\}}$, we restrict the model set to $M = 2^{\{1,2,3\}} \setminus \{\emptyset\}$, the thresholds coincide independently of the employed tiebreaking rule.} %
  Since models that induce a smaller posterior belief are associated with a higher expected fit in this case, less than fully informative communication does not relax the incentive-compatibility constraints. As with a fully informative strategy, the sender has incentives to deviate from the partially informative strategy associated with the weakest incentive-compatibility constraints if $b>\frac{1}{40}$, see Figure \ref{fig:ex2:ii} for an illustration.
 \end{enumerate}
 \begin{figure}[ht]
 \centering
    \begin{tikzpicture}[scale=50]
        \begin{scope}
        %%sigma scope
		%drawings
		\draw (0.17,0)--(0.42,0);

        %sigma, rho
        \node at (0.17,0.016) [left,shift={(-0.5,0)}]{$\sigma$:};
        \draw (0.17,0) node[left,shift={(-0.5,0)}]{$\rho$:};
  
        \def\blisspoints{1/5,1/4,1/3}
        \def\sigmapoints{1/2*1/5+1/2*1/4}
        \def\rhopoints{1/5,1/4}
		\foreach \x in \blisspoints{
		      \draw (\x,0)--(\x,-0.01)node[below,scale=0.75]{$\x$};
        }
        \draw (0.4,0)--(0.4,-0.01)node[below,scale=0.75]{$1/2$};
            \draw(11/30-0.002,-0.003)--(11/30,0.003);
            \draw(11/30+0.001,-0.003)--(11/30+0.003,0.003);

        \draw (1/5,-0.033)node[scale=0.75,align=center]{0 \faIcon{check-circle}\\3 \faIcon{times-circle}};
        \draw (1/4,-0.033)node[scale=0.75,align=center]{0 \faIcon{check-circle}\\2 \faIcon{times-circle}};
        \draw (1/3,-0.033)node[scale=0.75,align=center]{0 \faIcon{check-circle}\\1 \faIcon{times-circle}};
        \draw (0.4,-0.033)node[scale=0.75,align=center]{0 \faIcon{check-circle}\\0 \faIcon{times-circle}};

        \draw (0.25,0.04)node[]{$h=(0,0,0)$ $\hat{=}$  0 \faIcon{check-circle} 3 \faIcon{times-circle}};
        
        %boundaries
        %(i)sigma
        \foreach \s in \sigmapoints{
                \draw[thick,dashed] (\s,0.025)--(\s,-0.025);
        }
        %messages
        \draw[decoration={brace,raise=10pt},decorate] (1/5-1/150,0) -- node[above=14pt] {$r_{1}$} (1/5+1/150,0);
        \draw[decoration={brace,raise=10pt},decorate] (1/4-1/150,0) -- node[above=14pt] {$r_{2}$} (0.4+1/150,0);
        
        %rho
        \foreach \a in \rhopoints{
            \filldraw (\a,0) circle (0.06pt);
        \draw[decoration={amplitude=7pt,mirror,raise=8pt},decorate] (1/5,-.05) -- node[below=1pt] {$\Delta=\frac{1}{20}$} (1/4,-.05);
        \draw [-] (1/5,-.0475) to (1/5,-.0525);
\draw [-] (1/4,-.0475) to (1/4,-.0525);
        }
        \end{scope}
    \end{tikzpicture}
\caption{Partially informative strategy $\sigma$, where $M = 2^{\{1,2,3\}}$ and $h= (0,0,0)$ entails three failures (\faIcon{times-circle}) in Example \ref{ex2} \eqref{ex2:ii}. The optimal receiver response is indicated by the solid dots.}
\label{fig:ex2:ii}
\end{figure}
\end{example}

Example \ref{ex2} shows that the informativeness threshold that marks when fully informative communication is possible in equilibrium may vary largely with the history. We give a simple measure of explainability to identify which histories are more difficult to explain than others. To this end, we say that the sender can \emph{explain} history $h\in H$ if she can credibly reveal all models $m\in M$ under $h$, i.e.,
\begin{align}\label{eq: C_truthful}
U_{m^T,h}(a(m^T,h,0), b) \geq U_{m^T,h}(a^+(m^T,h,0),b) \text{ for all non-maximal } m^T \in M, %\tag{\mathcal C(h,b)} 
\end{align}
where $a^+(m^T,\tilde{h},0) = \min\{ a(m,\tilde{h},0) \mid m\in M,\ a(m,\tilde{h},0) > a(m^T,\tilde{h},0)\}$ for any $\tilde{h} \in H$. %
In other words, the sender can explain a history if the policy implications of distinct narratives---not the narratives themselves---are sufficiently different. %
We then say that history $h$ is \emph{more difficult to explain} than $h'$ if, for every bias $b$, the sender can explain $h'$ whenever she can explain $h$. By Proposition \ref{pro:fully_inf}, it follows readily that the sender can explain $h$ if and only if $b\leq \underline{b}(h)$. 
\begin{corollary}\label{Corollary: easier to explain histories}
    A history $h$ is more difficult to explain than $h'$ if and only if $\underline{b}(h) \leq \underline{b}(h')$.
\end{corollary}
Furthermore, under quadratic loss we have $\underline b(h)=\frac{1}{2}\min_{a,a' \in \{a(m,h,0) \mid m\in M\}, a \neq a'} |a-a'|$, and thus that a history is more difficult to explain than another if it comes with a smaller minimal narrative-induced action shift.

\begin{remark}
If $u_S(a,\theta,b) = -(\theta+b-a)^2$, history $h$ is more difficult to explain than $h'$ if and only if 
\begin{equation*}
    \min_{a,a' \in \{a(m,h,0) \mid m\in M\}, a \neq a'} |a-a'|\leq \min_{a,a' \in \{a(m,h',0) \mid m\in M\}, a \neq a'} |a-a'|.
\end{equation*}
\end{remark}

Thus, histories that are difficult to explain are exactly those under which some narratives translate into similar yet distinct policy consequences.

Finally, we investigate whether the sender prefers more informative equilibria. The following proposition indeed establishes that the sender prefers more informative equilibria---and thus a most informative one---if facing a receiver who applies the MLEU ambiguity rule. The result furthermore extends to the notion of Blackwell informativeness (cf. Remark \ref{Remark: more informative from equilibrium to Blackwell}).
\begin{proposition}\label{Proposition: MLEU equilibria prefer informativeness}
Let $(\sigma,\rho)$ and $(\sigma', \rho')$ be MLEU equilibria and
    \begin{enumerate}[(i)]
        \item $(\sigma,\rho)$ be weakly more informative than $(\sigma', \rho')$, or
        \item $\sigma$ be weakly more Blackwell informative than $\sigma'$.
    \end{enumerate}
    Then, the sender prefers the outcome of $(\sigma,\rho)$ over $(\sigma',\rho')$ for any $m^T$.
\end{proposition}
A more (Blackwell) informative equilibrium discriminates more finely between different models. Under the MLEU rule, splitting up an interval into subintervals means the sender can still induce the same actions as before (and an additional one), as the respective models will continue to maximize the expected fit across some interval. The sender thus could achieve the same outcome as before splitting up the interval, but has no incentives to do so by definition of equilibrium.\label{Proposition: MLEU equilibria prefer informativeness:discussion}

A similar statement is not true for the MEU or the Bayesian ambiguity rule. As the examples in Figure \ref{Figure: MEU less informative equilibrium better} show, splitting up an interval into subintervals does not necessarily yield a superset of actions and may thus be worse for the sender. %
The counterexample marks a stark contrast to Theorem 2 in \cite{colo2024communicating}, where the most (Blackwell) informative equilibrium is interim dominant under MEU preferences. %
In further contrast to \citeauthor{colo2024communicating}'s Theorem 1, not all `cut-off' models defining the interval partition lie below the action under full uncertainty: under $\sigma$ and either ambiguity rule from Figure \ref{Figure: MEU less informative equilibrium better}, the cut-off model $m=\{1,2,3\}$ induces an action $3/5=0.6$ strictly greater than the resp.\ babbling actions $\tfrac{13}{25} =0.52$ for MEU and $\tfrac{271}{480} \approx 0.56$ for the Bayesian rule.
\begin{figure}
    \centering
    \begin{tikzpicture}[scale=18]
        %%sigma scope
		%drawings
        
        \begin{scope}
        \draw (0.3,0)--(0.8,0);
        \draw (0.3,0.05)--(0.8,0.05);

        %sigma, rho
        \node at (0.3,0.086) [left,shift={(-0.5,0)}]{$\sigma$:};
        \draw (0.3,0.05) node[left,shift={(-0.5,0)}]{$\rho_{\MEU}$:};
        \draw (0.3,0) node[left,shift={(-0.5,0)}]{$\rho_{\Bayes}$:};
        
        \def\blisspoints{1/3,1/2,3/5,2/3,3/4}
        \def\sigmapoints{1/2*1/3+1/2*1/2,1/2*1/2+1/2*3/5}
        \def\rhopoints{1/3,1/2,2/3}
        \def\rhopointsB{1/3,1/2,161/240}
		\foreach \x in \blisspoints{
            \draw (\x,0.05)--(\x,0.035)node[below,scale=0.75]{$\x$};
        }
        \foreach \x in {1/3,1/2,3/5,3/4}{
                \draw (\x,0)--(\x,-0.015)node[below,scale=0.75]{$\x$};
            }
        \draw (2/3,0)--(2/3,-0.015);
        \draw (121/180,-0.015)node[below,scale=0.9]{$\frac{161}{240}$};

        \draw (1/3,-0.08)node[scale=0.75,align=center]{0 \faIcon{check-circle}\\1 \faIcon{times-circle}};
        \draw (1/2,-0.08)node[scale=0.75,align=center]{1 \faIcon{check-circle} \ 0 \faIcon{check-circle}\\1 \faIcon{times-circle} \ 0 \faIcon{times-circle}};
        \draw (3/5,-0.08)node[scale=0.75,align=center]{2 \faIcon{check-circle}\\1 \faIcon{times-circle}};
        \draw (2/3,-0.08)node[scale=0.75,align=center]{1 \faIcon{check-circle}\\0 \faIcon{times-circle}};
        \draw (3/4,-0.08)node[scale=0.75,align=center]{2 \faIcon{check-circle}\\0 \faIcon{times-circle}};

        \draw (0.55,0.15)node[]{$h=(1,0,1)$ $\hat{=}$  2 \faIcon{check-circle} 1 \faIcon{times-circle}};
        
        %boundaries
        %(i)sigma
        \foreach \s in \sigmapoints{
                \draw[thick,dashed] (\s,0.095)--(\s,-0.055);
        }
        %messages
        \draw[decoration={brace,raise=10pt},decorate] (1/3-1/60,0.05) -- node[above=14pt] {$r_{1}$} (1/3+1/60,0.05);
        \draw[decoration={brace,raise=10pt},decorate] (1/2-1/60,0.05) -- node[above=14pt] {$r_{2}$} (1/2+1/60,0.05);
        \draw[decoration={brace,raise=10pt},decorate] (3/5-1/60,0.05) -- node[above=14pt] {$r_{3}$} (3/4+1/60,0.05);

        %rho
        \foreach \a in \rhopoints{
            \filldraw (\a,0.05) circle (0.15pt);
        }
         %rho
        \foreach \a in \rhopointsB{
            \filldraw (\a,0) circle (0.15pt);
        }
        
        \end{scope}

        %%%second scope
        \begin{scope}[shift={(0,-0.24)}]
        \draw (0.3,0)--(0.8,0);
        \draw (0.3,0.05)--(0.8,0.05);

        %sigma, rho
        \node at (0.3,0.086) [left,shift={(-0.5,0)}]{$\sigma'$:};
        \draw (0.3,0.05) node[left,shift={(-0.5,0)}]{$\rho'_{\MEU}$:};
        \draw (0.3,0) node[left,shift={(-0.5,0)}]{$\rho'_{\Bayes}$:};
        
        \def\blisspoints{1/3,1/2,3/5,2/3,3/4}
        \def\sigmapoints{1/2*1/3+1/2*1/2}
        \def\rhopoints{1/3,8/15}
        \def\rhopointsB{1/3,251/420}
		\foreach \x in \blisspoints{
		      \draw (\x,0.05)--(\x,0.035)node[below,scale=0.75]{$\x$};
        }
        \foreach \x in {1/3,1/2,2/3,3/4}{
            \draw (\x,0)--(\x,-0.015)node[below,scale=0.75]{$\x$};
        }
        \draw (3/5,0)--(3/5,-0.01);

        \draw (1/3,-0.08)node[scale=0.75,align=center]{0 \faIcon{check-circle}\\1 \faIcon{times-circle}};
        \draw (1/2,-0.08)node[scale=0.75,align=center]{1 \faIcon{check-circle} \ 0 \faIcon{check-circle}\\1 \faIcon{times-circle} \ 0 \faIcon{times-circle}};
        \draw (3/5,-0.08)node[scale=0.75,align=center]{2 \faIcon{check-circle}\\1 \faIcon{times-circle}};
        \draw (2/3,-0.08)node[scale=0.75,align=center]{1 \faIcon{check-circle}\\0 \faIcon{times-circle}};
        \draw (3/4,-0.08)node[scale=0.75,align=center]{2 \faIcon{check-circle}\\0 \faIcon{times-circle}};
        
        %boundaries
        %(i)sigma
        \foreach \s in \sigmapoints{
                \draw[thick,dashed] (\s,0.095)--(\s,-0.055);
        }
        %messages
        \draw[decoration={brace,raise=10pt},decorate] (1/3-1/60,0.05) -- node[above=14pt] {$r_{1}$} (1/3+1/60,0.05);
        \draw[decoration={brace,raise=10pt},decorate] (1/2-1/60,0.05) -- node[above=14pt] {$r_{2}$} (3/4+1/60,0.05);

        %rho
        \foreach \a in \rhopoints{
            \filldraw (\a,0.05) circle (0.15pt);
        }
        \foreach \a in \rhopointsB{
            \filldraw (\a,0) circle (0.15pt);
        }
        \draw (8/15,0.045)node[below,scale=0.9]{$\frac{8}{15}$};
        \draw (251/420,-0.005)node[below,scale=0.9]{$\frac{251}{420}$};
        \end{scope}

    \end{tikzpicture}
    \caption{Equilibria in the setting of Section \ref{ex:uniform_random_quadratic} with $h=(1,0,1)$ and $b = 0.08$ for the MEU rule and the Bayesian rule ($\phi(x) = x$) with prior $\upsilon(m) := \tfrac{\Pr(h \mid m, F_0)^{-1}}{\sum_{m'\in M} \Pr(h \mid m', F_0)^{-1}}$. While $(\sigma,\rho)$ is more informative than $(\sigma',\rho')$, the sender prefers the MEU-equilibrium action $\tfrac{8}{15}\approx 0.53$ (resp.\ $\tfrac{251}{420} \approx 0.6$ for Bayes) under $\sigma'$ over $\tfrac{1}{2}$ under $\sigma$ for the true model $m^T=\{1,2\}$, since her bliss point is $a(m^T,h,b) = 0.58$.}
    \label{Figure: MEU less informative equilibrium better}
\end{figure}

\section{Analysis of the running example}\label{Section: conrete analysis Example}
In this section, we provide some more insights into the running example, characterizing the informativeness thresholds on the conflict of interest (cf.\ Proposition \ref{pro:fully_inf}). %
Recall from Section \ref{ex:uniform_random_quadratic} that the running example features a hawkish central bank adviser (sender) who communicates with a dovish policymaker (receiver) before an interest-rate decision.
Underlying medium-term inflationary pressure $\theta$ is drawn from $F_0 = \mathcal U(0,1)$, and the observable data is given by a sequence $h=(h_1,h_2,\ldots,h_K) \in \{0,1\}^K$ of public macroeconomic releases categorized as inflationary ($1$) or non-inflationary ($0$). The true model $m^T\in M=2^{\{1,2,\ldots,K\}}$ indicates the instances in which underlying inflationary pressure contributed to the observed release, i.e., $h_1,\dots,h_K$ are independent conditional on $\theta$, with
\begin{align*}
\Pr(h_k=1\ |\ \theta)=\left\{\begin{array}{cl} \theta, & \mbox{if }k\in m^T\\ \frac{1}{2}, & \mbox{else} \end{array}\right. \text{for all } k=1,2,\ldots, K.
\end{align*}
Furthermore, the adviser's payoff function is $u_S(a,\theta,b)=-(\theta+b-a)^2$, reflecting her preference for a tighter monetary policy at every level of underlying inflationary pressure.

Observe first that in our concrete setting $h^\Sigma := \sum_{k=1}^K h_k$ is a sufficient statistic for history $h$, counting the bare number of successes and failures. Second, the expected value given $m$ and $h$ is the bliss point of the receiver, 
\begin{align*}
a(m,h,0) = \E[\theta \mid m,h]=\frac{\sum_{i\in m}h_i+1}{\#m+2}.
\end{align*}

The biased sender chooses her strategy in order to induce a favorable posterior belief (about the underlying inflation pressure) of the receiver. We know from Proposition \ref{pro:fully_inf} that she is willing to reveal the true model to him if interests are sufficiently aligned, $b\leq\underline{b}(h^{\Sigma}) := \underline{b}(h)$, while she cannot credibly reveal information in a way that favors her if her bias is too large, $b>\overline{b}(h^{\Sigma}) := \overline{b}(h)$.
We establish that the former threshold concerning whether fully informative communication is feasible, $\underline{b}(h^{\Sigma})$, is not only independent of the ambiguity rule but also symmetric, and that extreme histories are the most difficult to explain, as $\underline b(h^\Sigma)$ attains its minimum at these histories (Corollary \ref{Corollary: easier to explain histories}).

\begin{proposition}\label{pro:extreme_histories}
$\underline b(h^\Sigma)$ is independent of the ambiguity rule, symmetric, i.e., $\underline b(h^\Sigma)=\underline b(K-h^\Sigma)$ for any $h^\Sigma$, and $\{0,K\}=\argmin_{h^\Sigma=0,1,\ldots,K}\underline b(h^\Sigma).$
\end{proposition}

Figure \ref{Figure: underline{b} not quasi-concave} illustrates our findings. Somewhat surprisingly:

\begin{remark}
$\underline b(h^\Sigma)$ generally is not quasi-concave.
\end{remark}

\begin{figure}[ht]
    \begin{subfigure}{0.25\textwidth}
    \begin{tikzpicture}[xscale=0.9,yscale=150]
    
    %% K=3

    %grid
    \draw[<->] (0,0.035)--(0,0.01)--(3.4,0.01)node[below,scale=0.75]{$h^{\Sigma}$};
    \foreach \p in {0,...,3} {
       \draw (\p,0.01)--(\p,0.009)node[below,scale=0.75]{\p};
   }
        \foreach \x in {0.01,0.02,0.03}{
        \draw (0,\x)--(-0.1,\x)node[left,scale=0.75]{\x};
    }
    
        %list of points to be connected
        \def\points{(0, 1/40), (1, 1/30), (2, 1/30), (3, 1/40)}   
        %draws lines between the points in order of occurence
        \foreach \p [count=\i,remember=\p as \lastp] in \points{
        \ifnum\i>1\relax
        \draw[blue] \lastp -- \p;
        \fi
}
    \end{tikzpicture}
    \end{subfigure}
    \hfill
    \begin{subfigure}{0.3\textwidth}
    \begin{tikzpicture}[xscale=0.85,yscale=150]

    %%K = 4
    %grid
    \draw[<->] (0,0.035)--(0,0.01)--(4.4,0.01)node[below,scale=0.75]{$h^{\Sigma}$};
    \foreach \p in {0,...,4} {
        \draw (\p,0.01)--(\p,0.009)node[below,scale=0.75]{\p};
    }
        \foreach \x in {0.01,0.02,0.03}{
        \draw (0,\x)--(-0.1,\x)node[left,scale=0.75]{\x};
    }

        %list of points to be connected
        \def\points{(0, 1/60), (1, 1/40), (2, 1/30), (3, 1/40), (4, 1/60)
}   
        %draws lines between the points in order of occurence
        \foreach \p [count=\i,remember=\p as \lastp] in \points{
        \ifnum\i>1\relax
        \draw[blue] \lastp -- \p;
        \fi
}
    \end{tikzpicture}
    \end{subfigure}
    \hfill
    \begin{subfigure}{0.35\textwidth}
    \begin{tikzpicture}[xscale=0.71,yscale=150]

    %%K = 5

        %grid
    \draw[<->] (0,0.035)--(0,0.01)--(5.4,0.01)node[below,scale=0.75]{$h^{\Sigma}$};
    \foreach \p in {0,...,5} {
        \draw (\p,0.01)--(\p,0.009)node[below,scale=0.75]{\p};
    }
        \foreach \x in {0.01,0.02,0.03}{
        \draw (0,\x)--(-0.1,\x)node[left,scale=0.75]{\x};
    }
    
        %list of points to be connected
        \def\points{(0, 1/84), (1, 1/60), (2, 1/70), (3, 1/70), (4, 1/60), (5, 1/84)}   
        %draws lines between the points in order of occurence
        \foreach \p [count=\i,remember=\p as \lastp] in \points{
        \ifnum\i>1\relax
        \draw[blue] \lastp -- \p;
        \fi
}
    \end{tikzpicture}
    \end{subfigure}
    \vskip5mm
    \centering
        \begin{subfigure}{1\textwidth}
    \begin{tikzpicture}[xscale=0.62,yscale=5500]
    %%K = 20

    %grid
    \draw[<->] (0,0.0016)--(0,0.001)--(21,0.001)node[below,scale=0.75]{$h^{\Sigma}$};
    \foreach \p in {0,...,20} {
        \draw (\p,0.001)--(\p,0.00097)node[below,scale=0.75]{\p};
    }
        \foreach \x in {0.001,0.0015}{
        \draw (0,\x)--(-0.1,\x)node[left,scale=0.75]{\x};
    }
    
        %list of points to be connected
        \def\points{(0, 1/924),
 (1, 1/840),
 (2, 1/760),
 (3, 1/684),
 (4, 1/714),
 (5, 1/714),
 (6, 1/836),
 (7, 1/680),
 (8, 1/748),
 (9, 1/798),
 (10, 1/798),
 (11, 1/798),
 (12, 1/748),
 (13, 1/680),
 (14, 1/836),
 (15, 1/714),
 (16, 1/714),
 (17, 1/684),
 (18, 1/760),
 (19, 1/840),
 (20, 1/924)
}   
        %draws lines between the points in order of occurence
        \foreach \p [count=\i,remember=\p as \lastp] in \points{
        \ifnum\i>1\relax
        \draw[blue] \lastp -- \p;
        \fi
}
    \end{tikzpicture}
    \end{subfigure}    
    \caption{$\underline b(h^\Sigma)$ as a function of $h^\Sigma$ for $K=3,4,5$ (top left to right) and $K=20$ (bottom).}
    \label{Figure: underline{b} not quasi-concave}
    %\label{Figure: b underline as function of h^Sigma}
\end{figure}
\noindent%
Second, we consider the threshold $\overline{b}(h^{\Sigma})$ concerning whether partially informative communication is feasible. We show that we no longer have symmetry in this case due to the receiver's residual uncertainty.

\begin{proposition}\label{prop:partial_asym}
Under the MLEU and MEU ambiguity rule, $\overline{b}(h^\Sigma)$ is asymmetric if $K\geq 2$. In the case of
\begin{enumerate}[(i)]
  \item MLEU,\footnote{We break the tie in expected fit between a narrative $m$ such that $\#m=1$ and one such that $\#m=0$ (both $2^{-K}$) by adopting the first one.} we have $\overline{b}(K)=\frac{K}{4(K+2)}>\overline{b}(0)=\underline b(0)=\frac{1}{2(K+1)(K+2)}$,
  \item MEU, we have $\overline{b}(K) = \max \{ \tfrac{1}{12}, \tfrac{K}{8(K+3)} \} < \overline{b}(0) = \tfrac{K+1}{6(K+2)}$.
\end{enumerate}
\end{proposition}

\noindent%
To understand the intuition, recall from Section \ref{Section: Informativeness} that we can characterize a $2$-step equilibrium by a threshold model $m^*$ such that $V(m^*,h,b)\leq 0$, i.e., the sender has no incentives to deviate if $m^T=m^*$. The proof establishes that for the MLEU and MEU ambiguity rule, the $2$-step partition which determines the threshold $\overline{b}(h^\Sigma)$ is such that the first partition element is a singleton solely containing the minimal model with respect to the order of receiver bliss points, denoted $m^*(h^\Sigma)$. %
Thus, $\overline{b}(h^\Sigma)$ solves $V(m^*(h^\Sigma),h,b)=0$ and depends on the induced actions, in particular the receiver's action $a(\{ m \mid m >m^*(h^\Sigma) \} ,h)$ upon a deviation at $m^T=m^*(h^\Sigma)$. Note that $a(\{ m \mid m >m^*(h^\Sigma) \} ,h)$ depends on the ambiguity rule and the concrete history $h^\Sigma$ through the order of models it induces. 
In particular, $m^*(0)\neq m^*(K)$ and thus $a(\{ m \mid m >m^*(0) \},h)\neq a(\{ m \mid m >m^*(K) \},h)$. 
The higher the action chosen by the receiver, the more loose the sender's incentive-compatibility constraints and, thus, the higher $\overline{b}(h^\Sigma)$. %
Under the MLEU ambiguity rule, this action depends on the expected fit:
\begin{example}[MLEU ambiguity rule]\label{Ex:running_upper_threshold_MLEU} Suppose that $K=3$, $b\in (\frac{1}{15},\frac{3}{20}]$, and that the receiver behaves according to the MLEU ambiguity rule. As can be seen from Figure \ref{fig:running_overline b_MLEU}, informative communication is feasible in this parameter range for $h^\Sigma=3$ but not for $h^\Sigma=0$:
\begin{enumerate}[(i)]
  \item $h^\Sigma=3$. Any most informative equilibrium is partially informative, e.g., the one with $r=r_1$ if $m^T=\emptyset$, i.e., if no data point is relevant, and $r=r_2$ else. Note that a model's expected fit increases in the posterior it induces. After receiving report $r=r_2$, models that induce posteriors $\frac{2}{3}$, $\frac{3}{4}$, and $\frac{4}{5}$ are feasible, such that the receiver takes action $\frac{4}{5}$. Since $b\leq \frac{3}{20}$, the sender prefers action $a(m^T,h,0)=\frac{1}{2}$ over $\frac{4}{5}$ if $m^T=\emptyset$.      
 
  \item $h^\Sigma=0$. Any most informative equilibrium is uninformative. In this case, a model's expected fit decreases in the posterior it induces, such that any partially informative communication strategy is associated with weakly tighter incentive-compatibility constraints than a fully informative strategy. 
\end{enumerate}
\end{example}

Under the MEU ambiguity rule, the action $a(\{ m \mid m >m^*(h^\Sigma) \} ,h)$ upon a deviation at $m^T=m^*(h^\Sigma)$ depends on the variance associated with models $m^T>m^*(h^\Sigma)$:

\begin{example}[MEU ambiguity rule]\label{Ex:running_upper_threshold_MEU} Suppose that $K=3$, $b\in (\frac{1}{8},\frac{2}{15}]$, and that the receiver behaves according to the MEU ambiguity rule. As can be seen from Figure \ref{fig:running_overline b_MEU}, informative communication is feasible in this parameter range for $h^\Sigma=0$ but not for $h^\Sigma=3$:
\begin{enumerate}[(i)]
  \item $h^\Sigma=0$. Any most informative equilibrium is partially informative, e.g., the one with $r=r_1$ if $m^T=\{1,2,3\}$, i.e., if all data points are relevant, and $r=r_2$ else. After receiving report $r=r_2$, models that induce posteriors $\frac{1}{4}$, $\frac{1}{3}$, and $\frac{1}{2}$ are feasible. Note that a higher posterior is associated with a higher variance, such that the receiver's hedging yields action $\frac{7}{15}$, close to $\frac{1}{2}$. Since $b\leq \frac{2}{15}$, the sender prefers action $a(m^T,h,0)=\frac{1}{5}$ over $\frac{7}{15}$ if $m^T=\{1,2,3\}$.      
 
  \item $h^\Sigma=3$. Any most informative equilibrium is uninformative. Note that the partially informative strategy associated with the weakest incentive-compatibility constraints is such that $r=r_1$ if $m^T=\emptyset$, i.e., if no data point is relevant, and $r=r_2$ else. After receiving report $r=r_2$, models that induce posteriors $\frac{2}{3}$, $\frac{3}{4}$, and $\frac{4}{5}$ are feasible. Note that a higher posterior is associated with a lower variance, such that the receiver's hedging yields action $\frac{2}{3}$ equal to the lowest posterior induced by some feasible model. As a result, informative communication is precluded: since $b> \frac{1}{8}$, the sender prefers action $\frac{2}{3}$ over $a(m^T,h,0)=\frac{1}{2}$ if $m^T=\emptyset$.
\end{enumerate}
\end{example}

\begin{figure}[ht]
    \begin{center}
    \begin{subfigure}{0.4\textwidth}
    \begin{tikzpicture}[xscale=1.2,yscale=25]
    
    %% K=3, MLEU

    %grid
    \draw[<->] (0,0.22)node[left,scale=0.75]{$\overline{b}(h^\Sigma)$}--(0,0)--(3.4,0)node[below,scale=0.75]{$h^{\Sigma}$};
    \foreach \p in {0,...,3} {
       \draw (\p,0)--(\p,-0.01)node[below,scale=0.75]{\p};
   }
        \foreach \x in {0,0.04,0.08,0.12,0.16,0.2}{
        \draw (0,\x)--(-0.1,\x)node[left,scale=0.75]{\x};
    }
    
        %list of points to be connected
        \def\points{(0, 1/40), (1, 5/24), (2, 5/24), (3, 3/20)} 
        
        %draws lines between the points in order of occurence
        \foreach \p [count=\i,remember=\p as \lastp] in \points{
        \ifnum\i>1\relax
        \draw[blue] \lastp -- \p;
        \fi
}

    %grid for non-babbling
    \draw[densely dotted] (0,1/15)--(3.2,1/15)node[right,scale=0.75]{$\frac{1}{15}$};
    \draw[densely dotted] (0,3/20)--(3.2,3/20)node[right,scale=0.75]{$\frac{3}{20}$};
    
    \end{tikzpicture}
    \caption{MLEU rule}
    \label{fig:running_overline b_MLEU}
    \end{subfigure}
    \quad
    \begin{subfigure}{0.4\textwidth}
    \begin{tikzpicture}[xscale=1.2,yscale=78]

    %%K = 3, MEU
    %grid
    \draw[<->] (0,0.15)node[left,scale=0.75]{$\overline{b}(h^\Sigma)$}--(0,0.08)--(3.4,0.08)node[below,scale=0.75]{$h^{\Sigma}$};
    \foreach \p in {0,...,3} {
        \draw (\p,0.08)--(\p,0.0768)node[below,scale=0.75]{\p};
   }
        \foreach \x in {0.08,0.10,0.12,0.14}{
        \draw (0,\x)--(-0.1,\x)node[left,scale=0.75]{\x};
    }

        %list of points to be connected
        \def\points{(0, 2/15), (1, 1/8), (2, 1/10), (3, 1/12)}   
        %draws lines between the points in order of occurence
        \foreach \p [count=\i,remember=\p as \lastp] in \points{
        \ifnum\i>1\relax
        \draw[blue] \lastp -- \p;
        \fi
        
}

  %grid for non-babbling
    \draw[densely dotted] (0,1/8)--(3.2,1/8)node[right,scale=0.75]{$\frac{1}{8}$};
    \draw[densely dotted] (0,2/15)--(3.2,2/15)node[right,scale=0.75]{$\frac{2}{15}$};
    
    \end{tikzpicture}
    \caption{MEU rule}
     \label{fig:running_overline b_MEU}
    \end{subfigure}
    \end{center}
    \caption{$\overline{b}(h^\Sigma)$ as a function of $h^\Sigma$ for $K=3$ for the MLEU and the MEU ambiguity rule. The dotted lines mark the thresholds on the bias in Example \ref{Ex:running_upper_threshold_MLEU} and \ref{Ex:running_upper_threshold_MEU}.}
    \label{fig:running_overline b}
\end{figure}

\section{Comparison with a \texorpdfstring{na\"ive}{naive} receiver}\label{Section: comparison naive}

In this section, we compare the MLEU equilibria in our framework with \citetalias{schwartzstein2021using}, where the receiver is na\"ive in the sense that he does not take the sender's strategic incentives into account. Instead, the receiver takes the sender's message $r=m$ at face value and adopts the proposed model $m$ if it has a higher expected fit than a default model $m^d$ given the observed history $h$, i.e., $\Pr(h|m,F_0)\geq \Pr(h|m^d,F_0)$. Otherwise he sticks to the default model. 
Formally, let $\Report=M$. The \emph{sender-optimal profile} $(\sigma^\dagger,\rho^\dagger)$ then satisfies
\begin{enumerate}[(i)]
    \item\label{naive:sender-optimal:i} $\sigma^\dagger(m^T) \in \argmax_{r} U_{m^T,h}(\rho^\dagger(r),b)$ for all $m^T\in M$,
    \item\label{naive:sender-optimal:ii} $\rho^\dagger(r) = \argmax_{a} U_{\tilde m,h}(a)$, where $\tilde m\in \argmax_{m\in \{r,m^d\}}\Pr(h \mid m, F_0)$, for all $r \in \Report$.\footnote{Like \citetalias{schwartzstein2021using}, we break ties in expected fit between the proposed model $r$ and the default $m^d$ in favor of the proposed model, i.e., $\tilde m=r$ if $\Pr(h \mid r, F_0)=\Pr(h \mid m^d, F_0)$.}
\end{enumerate}
First, note that the receiver does not take into account the sender's strategy. Second, the restriction of the comparison to MLEU equilibria appears natural in light of \eqref{naive:sender-optimal:ii}. 

We henceforth refer to the model of \citetalias{schwartzstein2021using} as the \emph{na\"ive model} and adopt their assumption (from Section III) that the receiver's default model is the true model, $m^d=m^T$. This choice is reasonable for at least two reasons: First, as \citetalias{schwartzstein2021using} point out, it allows to ask when the wrong story wins. Second, it can be seen as a `best case' scenario, under which the truthful communication benchmark has the same model-action profile as the truthful communication benchmark in our framework.\footnote{Note that the only other choice under which this is the case is a model with the lowest expected fit $m^d\in\argmin_{m\in M}\Pr(h|m,F_0)$, which however depends on the observed data $h$.}

A weak measure of the persuasive power of the sender is the set of models $m^T$ for which she prefers to induce an action $a(m^*,h,0)$, associated to another narrative $m^*\neq m^T$, over $a(m^T,h,0)$ and succeeds to do so under a profile $(\sigma,\rho)$ given the history $h$:
\begin{align*}
    \Meq &:= \left\{ m^T \in M \mid U_{m^T,h}(\rho(\sigma(m^T)),b) > U_{m^T,h}(a(m^T,h,0),b) \right\}.
\end{align*}
Take the na\"ive model, then $\Mna$ contains all models $m^T$ under which the sender wants to and successfully can convince a na\"ive receiver to move away from the action he would have chosen by default. She does so although the default model corresponds to the true model $m^T$, i.e., the wrong story wins. Likewise, for any MLEU equilibrium $(\sigma,\rho)$, $\Meq$ contains all models $m^T$ under which the sender is able to make the receiver take an action that she prefers over the one the receiver would like to take if he knew $m^T$. In other words, these sets contain the models under which the sender does better than under truthful communication. 
The following lemma proves that this will more often be the case facing a na\"ive receiver.

\begin{lemma}\label{Lemma: naive easier to convince}
For any MLEU equilibrium $(\sigma,\rho)$, we have $\Meq \subseteq \Mna$.
\end{lemma}

The key observation is that
\begin{align}\label{persuasive_power}
\begin{split}
&U_{m^T,h}(\rho(\sigma(m^T)), b)\\
\leq\, &\max \big\{ U_{m^T,h}(a(m^*,h,0), b) \mid m^*\in M \text{ with } \Pr(h \mid m^*,F_0) \geq \Pr(h \mid m^T,F_0) \big\}\\
=\, &U_{m^T,h}(\rho^\dagger(\sigma^\dagger(m^T)), b)
\end{split}
\end{align}
for every true model $m^T \in M$, since we consider the MLEU ambiguity rule. Thus, if the sender manages to profitably delude a receiver in equilibrium $(\sigma,\rho)$ into taking an action favorable to her, i.e., the left-hand side of Expression \eqref{persuasive_power} exceeds $U_{m^T,h}(a(m^T,h,0),b)$ (and thus $m^T\in \Meq$), then she can also do so when facing a na\"ive receiver. %
Moreover, Expression \eqref{persuasive_power} shows that the sender's expected utility is weakly \emph{larger} in the latter case, which gives rise to an arguably strong notion of persuasive power in the spirit of state-wise dominance:
\begin{definition}[Persuasive power ordering]\label{Def:persuasive_power}
The sender has \emph{(strictly) more persuasive power} under profile $(\sigma,\rho)$ than under $(\sigma',\rho')$ if $U_{m^T,h}(\rho(\sigma(m^T)), b)\geq U_{m^T,h}(\rho'(\sigma'(m^T)), b)$ for all $m^T \in M$ (and with strict inequality for some $m^T$). 
\end{definition}
At first glance, the notion of persuasive power provided in Definition \ref{Def:persuasive_power} may appear demanding: we cannot rank two profiles if the sender prefers one profile to the other under a model $m$ and vice versa under another model $m'\neq m$. However, as described above, it nonetheless allows for a comparison of the MLEU equilibria in our framework and the na\"ive model, with persuasive power being greater in the latter. Moreover, it follows from Lemma \ref{Lemma: naive easier to convince} and Expression \eqref{persuasive_power} that the ranking is strict if the inclusion relation is strict.
\begin{proposition}\label{Proposoition: naive easier to convince}
For any MLEU equilibrium $(\sigma,\rho)$, the sender has (strictly) more persuasive power in $(\sigma^\dagger,\rho^\dagger)$ than in $(\sigma,\rho)$ (if $\Meq \subsetneq \Mna$).
\end{proposition}

The following proposition characterizes cases in which the sender has the same or strictly more persuasive power when facing a na\"ive receiver.

\begin{proposition}\label{Proposition: generic indifference, improve upon naive}
For any history $h$ and MLEU equilibrium $(\sigma,\rho)$, the following two statements hold.
\begin{enumerate}[(i)]
    \item $\Meq = \Mna = \emptyset$ and the sender's expected payoff in $(\sigma,\rho)$ and in $(\sigma^\dagger,\rho^\dagger)$ is identical if $b \leq \underline{b}(h)$ and $(\sigma,\rho)$ is fully informative.
    \item $\Meq \subsetneq \Mna$ and the sender has strictly more persuasive power in $(\sigma^\dagger,\rho^\dagger)$ than in $(\sigma,\rho)$ if $b > \overline{b}(h)$, provided $U_{m,h}(a,b)$ allows for large conflicts of interest and there is $m^T,m^*$ with $\Pr(h \mid m^*,F_0) \geq \Pr(h \mid m^T, F_0)$ and $a(M,h) \leq a(m^T,h,0) < a(m^*,h,0)$.
\end{enumerate}
\end{proposition}
The intuition of the above result is as follows. On the one hand, if $b \leq \underline{b}(h)$, the best the sender can do is to reveal the true model and make the receiver take action $a(m^T,h,0)$. On the other hand, if $b> \overline{b}(h)$, provided preferences allow for large conflicts of interest, the only equilibrium is babbling, inducing the pooling action $a(M,h)$, i.e., the action the receiver would opt for if no communication took place. If a true model's best reply $a(m^T,h,0)$ exceeds the pooling action and the sender can credibly convince a na\"ive receiver to adopt a higher action, she strictly improves her expected utility in a case where equilibrium play cannot.

We conclude this section by providing an instance within our running example showing that the sender may also have strictly more persuasive power when facing a na\"ive receiver in case equilibrium play is informative. 

\begin{example}\label{Example: sender strictly prefers naive receiver}
    Consider the setting of Section \ref{ex:uniform_random_quadratic}, where $F_0 = \mathcal U(0,1)$, $u_S(a,\theta,b)=-(\theta+b-a)^2$, and $h$ is generated according to \eqref{ex:uniform_random_quadratic:tag1}, and the MLEU ambiguity rule. Suppose that $K = 3$, $h^{\Sigma} = 2$, $M = 2^{\{1,2,3\}} \setminus \{ \emptyset\}$ and $b \in (\tfrac{1}{20},\tfrac{3}{40}]\subseteq (\underline{b}(h),\overline{b}(h))$. A most informative MLEU equilibrium $(\sigma,\rho)$ is given by $\sigma(m)=r_1$ if $a(m,h,0) = \tfrac{1}{3}$, $\sigma(m) = r_2$ if $a(m,h,0) = \tfrac{1}{2}$ and $\sigma(m) = r_3$ if $a(m,h,0) \in \{ \tfrac{3}{5}, \tfrac{2}{3}, \tfrac{3}{4} \}$ (inducing posterior and action $\tfrac{3}{4}$). Facing a na\"ive receiver, the sender wants to and can use a narrative inducing a higher than the receiver's default action if $a(m^T,h,0) \in \{\tfrac{1}{2}, \tfrac{3}{5}, \tfrac{2}{3} \}$, see Figure \ref{Figure: naive vs strategic} for an illustration.
    
    We have $\Meq = \{ m \mid a(m,h,0)=\tfrac{2}{3} \}$, a proper subset of $\Mna = \{ m \mid a(m,h,0) \in \{ \tfrac{1}{2}, \tfrac{3}{5}, \tfrac{2}{3} \} \}$.
    Furthermore, if $a(m^T,h,0) = \tfrac{1}{2}$ and $a(m^T,h,0) = \tfrac{3}{5}$, the sender prefers to induce action $\tfrac{3}{5}$ over action $\tfrac{1}{2}$ and action $\tfrac{2}{3}$ over action $\tfrac{3}{4}$, respectively. She is thus  strictly better off facing a na\"ive receiver than facing equilibrium play. Finally, note that if $a(m^T,h,0) = \tfrac{3}{5}$, the sender would actually even prefer truthful communication over equilibrium play.
\begin{figure}
    \centering
    \begin{tikzpicture}[scale=18]
    %anchors
    	\draw (0.1,0);
    	\draw (0.85,0);
    %

%(i)
\draw (0.2,0);
\draw(0.25,0)node[left,shift={(-0.2,0)},scale=0.75,align=center]{facing MLEU\\equilibrium play}--(0.8,0);
\draw[->] (0.81,-0.038)node[right,align=left,scale=0.85]{$a(m,h,0)$} -- (0.785,-0.038);
\foreach \x in {1/3,1/2,3/5,2/3,3/4}
   \draw (\x,0.02)--(\x,-0.02)node[below]{$\x$};

%selfarrow
\foreach \x in {1/3,3/4}{
   \draw (\x,0.02)node[shift={(0,0.5)},thick,rotate=-150,scale=2]{$\circlearrowright$};
   }
   
\draw (1/2,0.02)node[shift={(0,0.5)},thick,rotate=-150,scale=2]{$\circlearrowright$};
%others
\draw[->] (3/5,0.025)to[out=35,in=100+45](3/4,0.025);
\draw[->] (2/3,0.025)to[out=35,in=100+45](3/4,0.025);

\begin{scope}[shift={(0,-0.15)}]
	%anchors
    	\draw (0.1,0);
    	\draw (0.85,0);
    %
	%(ii)
	\draw (0.2,0);
	\draw (0.25,0)node[left,shift={(-0.2,0)},scale=0.75,align=center]{facing a\\na\"ive receiver}--(0.8,0);
    \draw[->] (0.81,-0.038)node[right,align=left,scale=0.85]{$a(m,h,0)$} -- (0.785,-0.038);
	\foreach \x in {1/3,1/2,3/5,2/3,3/4}
   	\draw (\x,0.02)--(\x,-0.02)node[below]{$\x$};

	%selfarrow
	\foreach \x in {1/3,3/4}
   	\draw (\x,0.02)node[shift={(0,0.5)},thick,rotate=-150,scale=2]{$\circlearrowright$};
	%others
	
\draw[->] (1/2,0.025)to[out=35,in=100+45](3/5,0.025);

\draw[->] (3/5,0.025)to[out=35,in=100+45](2/3,0.025);

	\draw[->] (2/3,0.025)to[out=35,in=100+45](3/4,0.025);

	%expected fit
        \draw (1/3,-0.075)node[scale=0.85,align=center]{$\frac{1}{8}$};
        %\draw (1/2,-0.12)node[scale=0.85,align=center]{$\frac{1}{12}$ \ \ $\frac{1}{8}$};
        \draw (1/2,-0.075)node[scale=0.85,align=center]{$\frac{1}{12}$};
        \draw (3/5,-0.075)node[scale=0.85,align=center]{$\frac{1}{12}$};
        \draw (2/3,-0.075)node[scale=0.85,align=center]{$\frac{1}{8}$};
        \draw (3/4,-0.075)node[scale=0.85,align=center]{$\frac{1}{6}$};
        \draw[->] (0.81,-0.075)node[right,align=left,scale=0.85]{expected fit} -- (0.785,-0.075);
	
	%anchor, distance to caption
	\draw (0.5,-0.06);
\end{scope}
\end{tikzpicture}
    \caption{Illustration of equilibrium play and the na\"ive model in Example \ref{Example: sender strictly prefers naive receiver}. The arrows indicate the action the sender manages to make the receiver take given the action optimal for the receiver under the true model.
    }
    \label{Figure: naive vs strategic}
\end{figure}
\end{example}

\noindent%
Our results and examples show that the sender generally prefers to deal with a na\"ive receiver. One might suspect that the receiver is always better off when following equilibrium play. However, this is not true as can be seen from Figure \ref{Figure: naive vs strategic}: On the one hand, if $a(m^T,h,0) = \tfrac{1}{2}$, the sender reveals the true model under equilibrium play, while she can induce action $\tfrac{3}{5}$ if the receiver was na\"ive. On the other hand, if $a(m^T,h,0) = \tfrac{3}{5}$, equilibrium play enforces the action $\tfrac{3}{4}$, whereas both would prefer the action $\tfrac{2}{3}$ which were induced if the receiver was na\"ive.
\begin{remark}
The receiver's expected payoff may be strictly higher in the na\"ive model than following equilibrium play, or vice versa, depending on $m^T$. Hence, it is ex ante uncertain whether it is better for the receiver to be rational or na\"ive.
\end{remark}

\section{Conclusion}\label{Section: Conclusion}

%what we did
%framework
In this article, we model the communication of narratives as a cheap-talk game under model uncertainty. %
Our starting point was the literature initiated by \citetalias{schwartzstein2021using}, which formalizes narratives as likelihood functions explaining observable data in the presence of model uncertainty. A key assumption in this literature is a form of receiver na\"ivit\'e in the sense that the receiver does not take the sender's strategic incentives into account. %
In contrast, we provide a general game-theoretic framework in which the receiver is sophisticated and study equilibria under model uncertainty. To this end, we resolve any remaining model uncertainty by applying the novel concept of ambiguity rules, which allow for the study of MLEU, MEU, and smooth model preferences, which include the classical Bayesian approach.

%results
We characterize the set of equilibria by means of an algorithm: There is a positive integer $N$ such that there is an $n$-step equilibrium which divides the set of narratives into $n$ consecutive intervals if and only if $1\leq n \leq N$. %
Equilibrium communication pools narratives not according to the similarity of their likelihood functions but according to the similarity of the policy implications associated with them. %
We then show that the maximum number $N$ of consecutive intervals is decreasing in the conflict of interest and characterize informativeness thresholds that mark when full disclosure or partial disclosure is possible in equilibrium, respectively. Furthermore, we find that histories that are difficult to explain are those under which some narratives translate into similar yet distinct policy consequences. %
Finally, we compare our equilibrium framework under MLEU with \citetalias{schwartzstein2021using}. We establish that the sender has more persuasive power in a strong sense -- namely state-wise dominance -- facing a na\"ive receiver than facing a sophisticated receiver of our framework. Interestingly, the receiver does not generally prefer to be rational himself. 

\paragraph{Modeling assumptions.} 

Being based on \citetalias{schwartzstein2021using}, our framework diverges in important ways from the classical literature on cheap-talk communication following the seminal work by \cite{crawford1982strategic}: First, communication is not directly about the (payoff-relevant) state but about the model linking publicly observed outcomes and the state. %
Second, the model space is discrete. Third, the receiver faces ambiguity about the true model. 

At least to us, it was thus not clear a priori whether well-known results from the literature carry over to our framework. %
Most importantly, there is our equilibrium characterization, which closely resembles the one in \cite{crawford1982strategic}. Moreover, we obtain precise informativeness thresholds like in other cheap-talk models with an imperfectly informed  sender \citep[e.g.,][]{argenziano2016strategic,foerster2023strategic}. The relationship to the literature is less straightforward for the last set of results, as there is no obvious analogue to a na\"ive receiver \`a la \citetalias{schwartzstein2021using} in the setting of \cite{crawford1982strategic}. %
However, we could consider a receiver who takes the sender's message at face value. Such a receiver then follows the sender's recommendation, with the same outcome as under full delegation. While the sender clearly prefers delegation to cheap-talk communication, \cite{dessein2002authority} shows that the receiver also does so if the conflict of interest is small. On the contrary, for the receiver to prefer to be na\"ive in our framework not only requires the conflict of interest to be large enough to preclude truthful communication, but it also depends on the underlying true model. 

Finally, one may wonder whether a typical equilibrium selection criterion, such as the NITS (no incentive to separate) criterion by \cite{chen2008selecting}, has bite in our setting. The answer is yes and no. On the one hand, one can show that if an $n$-step equilibrium does not satisfy NITS, then there is an $n+1$-step equilibrium, implying that an $N$-step equilibrium satisfies NITS. On the other hand, even a babbling equilibrium may satisfy NITS due to the model space being finite.

\paragraph{Outlook.}

In light of \citetalias{schwartzstein2021using}, one might ask whether individuals take into account a sender's strategic incentives when being exposed to a narrative, as we assume they do,
 or simply adopt a narrative whenever it fits the data well, as they do. To this end, we conduct a laboratory experiment in a companion project \citep{bauch2026strategic}. Different from \cite{barron2024narrative}, we focus on whether subject receivers take the size of the conflict of interest between them and the subject sender into account when evaluating the communicated narrative. 

Our framework may also be extended in several directions. First, it does not embed the framework of \cite{crawford1982strategic} due to the finite model space, which precludes the sender from perfectly learning the state. %
One can show that this is possible in a tailor-made extension of our model to a specific continuous model space $M$. Second, one could introduce multiple senders who compete for convincing the receiver (or perhaps the receivers) along the lines of \cite{krishna2001model}. \citetalias{schwartzstein2021using} show that competition between persuaders pushes them to propose models that overfit the data. In turn, the receiver underreacts to the data. %
In our framework with a sophisticated receiver, we suspect competition to benefit the receiver similarly to \cite{krishna2001model}, with the extent depending on the exact protocol. %
Third, one could change the order of events, such that the data realizes only after the sender has communicated her narrative, similarly to \cite{aina2023tailored}. From the sender's point of view, the receiver's response to her narrative then is a random variable. Fourth, one could investigate alternatives to the NITS criterion in our setting. We leave these and related questions for future research.

%=============================================
% Bibliography
\setlength{\bibsep}{0pt}
\bibliography{references}
%=============================================

\newpage
\appendix
\section{Proofs}

\begin{proof}[Proof of Lemma \ref{Lemma: reduction lemma}]
	\begin{enumerate}[(i)]        
        \item Fix any induced action $a^* \in \rho'(\sigma'(M))$. Define $M^* := \{ m \in M \mid \rho'(\sigma'(m)) = a^* \}$, which is non-empty and contains some element $m^*$. Set $r^* := \sigma'(m^*)$ and define the sender strategy $\sigma(m) := r^*$ if $m \in M^*$ and $\sigma(m) := \sigma'(m)$ otherwise. Let $\rho$ be the receiver's best reply to $\sigma$. Then
        \begin{equation*}
            \rho(r^*) = a(M^*,h,0) \in \conv (\{\underbrace{\rho'(\sigma'(m))}_{= a^*}  \mid m \in M^* \}) = \{ a^* \},
        \end{equation*}
        i.e., $\rho(r^*) = a^* =  \rho'(r^*)$ and $\rho(r) = a(\sigma'^{-1}(r),h,0) = \rho'(r)$ if $r \in \sigma(M) \setminus \{ r^*\}$. By construction, $\rho \circ \sigma = \rho' \circ \sigma'$ and thus $\sigma$ is also a best reply to $\rho$. Note that the action $a^*$ is induced solely by the report $r^*$. Repeating this construction for all other induced actions yields the result.
        
		\item Let $a_1 < \dots < a_N$ be the distinct actions induced in equilibrium. By \eqref{item: reduction lemma rho injective} we can pass to the corresponding reduced equilibrium $(\sigma,\rho)$ and thus assume that the sender solely uses messages $r_1, \dots, r_N$ with $\rho(r_i) = a_i$. Let $M_i := \sigma^{-1}(r_i) \subseteq M$. We claim that $M_i$ is an interval. To this end, let $m<m'<m''$ with $m,m'' \in M_i$. For any other equilibrium action $a_j < a_i$, strict single crossing yields $U_{m',h}(a_i,b)>U_{m',h}(a_j,b)$ since $U_{m,h}(a_i,b)\geq U_{m,h}(a_j,b)$, as sending $r_i$ is optimal under $m$. Assume by means of contradiction that inducing an equilibrium action $a_j > a_i$ is optimal under $m'$, i.e., $U_{m',h}(a_j,b) \geq U_{m',h}(a_i,b)$. By strict single crossing, we thus find $U_{m'',h}(a_j,b)>U_{m'',h}(a_i,b)$ in violation of sending $r_i$ being optimal under $m''$. Consequently, the unique optimal equilibrium action the sender can induce under $m'$ is $a_i$, i.e., $m' \in M_i$. %
        The $M_i$ form a partition of $M$, and since $a_i = \rho(r_i) = a(M_i,h) \in \conv(\{ a(m,h,0) \mid m \in M_i \})$ we find $m_i \leq m_j$, i.e., $a(m_i,h,0) \leq a(m_j,h,0)$, whenever $i<j$ and $m_i \in M_i, m_j \in M_j$. In other words, the $M_i$ form a consecutive partition and $a_i = \rho(\sigma(m_i)) < \rho(\sigma(m_j)) = a_j$ implies $a(m_i,h,0) \leq a(m_j,h,0)$. Consistent ordering concludes the proof.\qedhere
\end{enumerate}
\end{proof}

\begin{proof}[Proof of Corollary \ref{Corollary: upper bound N stage}]
Let $a_1 < \dots < a_N$ be the distinct actions induced in equilibrium $(\sigma,\rho)$. By Lemma \ref{Lemma: reduction lemma} \eqref{item: reduction lemma ordering}, we can assume $\rho(r_i) = a_i$ and let $M_i := \sigma^{-1}(r_i) \subseteq M$ for messages $r_1, \ldots, r_N \in \Report$. Moreover, we have $a_i = a(M_i,h)\leq \bar a_i:= \max_{m\in M_i} a(m,h,0)$. %
Assume by means of contradiction that $\bar{a}_i \geq a_{i+1}$ and let $m_i \in M_i$ fulfill $\bar{a_i} = a(m_i,h,0)$. Then, $a_i <a_{i+1} \leq \bar{a_i} =a(m_i,h,0) < a(m_i,h,b)$. By concavity of the sender's utility function, she thus has a profitable deviation by sending $r_{i+1}$ instead of $r_i$ if facing $m_i$. Consequently, $\bar a_i< a_{i+1}$. Hence, $a_1 \leq \bar a_1 < a_2\leq \bar a_2< \dots < a_N\leq \bar a_N$. In particular, the $\bar a_i$ are $N$ distinct receiver bliss points, which yields $\#(\rho\circ \sigma)(M)\leq \#\{a(m,h,0) \mid m\in M\}$.
\end{proof}

\begin{proof}[Proof of Theorem \ref{Theorem: N step equilibrium}]
Fix any history $h$ and bias $b$ in the following. By Corollary \ref{Corollary: upper bound N stage}, an equilibrium induces at most $\#\{a(m,h,0) \mid m\in M\}$ different actions. Consequently, there is a maximum number $N\leq \#\{a(m,h,0) \mid m\in M\}$ such that an $N$-step equilibrium exists. % 
For any $n$-step equilibrium, the corresponding reduced equilibrium $(\sigma,\rho)$ uses exactly $n$ different messages $r_1, \ldots, r_n$ without changing the model-action profile $(m,(\rho \circ \sigma)(m))$ by the Reduction Lemma \ref{Lemma: reduction lemma}. The proof of the Reduction Lemma furthermore establishes that $M_i = \sigma^{-1}(r_i)$ is a consecutive partition of $M$ with $\sigma(m) = r_i$ for all $m \in M_i$ and the equilibrium conditions.

It remains to show that for every $1 \leq n \leq N$ there exists an $n$-step equilibrium. We do so by means of an algorithm that constructs an $n$-step equilibrium out of an ($n+1$)-step equilibrium. Starting from any $N$-step equilibrium, the claim follows. In the following, we explain the algorithm and show its properties.

Let $(\sigma,\rho)$ be an ($n+1$)-step equilibrium. If $n=1$ there is nothing to prove, as a babbling equilibrium always exists. Assume thus $n \geq 2$. % 
By passing to the corresponding reduced equilibrium $(\sigma,\rho)$, we find a consecutive partition $\{M_1,M_2,\ldots,M_{n+1}\}$ and messages $r_1,\ldots, r_{n+1}$ such that $\sigma(m) = r_i$ for all $m \in M_i$ and $\rho(r_i) < \rho(r_{i+1})$ for $i=1,\ldots,n$. %
Define the new strategy $\sigma^1$ resulting from $\sigma$ by merging the messages $r_n, r_{n+1}$. The new partition elements are thus $M^1_i = M_i$ for $1 \leq i \leq n-1$ and $M^1_n = M_n \cup M_{n+1}$, and we set $\sigma^1(m) = r_i$ whenever $m \in M^1_i$ for $1 \leq i \leq n$. %
If $\sigma^1$ together with its unique best reply $\rho^1$ of the receiver is an $n$-step equilibrium, we are done. %
Suppose now that $(\sigma^1,\rho^1)$ is not an equilibrium. Then, there is a model $m'$ for which the sender has a profitable deviation by sending a report different from $\sigma^1(m')$. Now, note that $\rho(r_n) \leq \rho^1(r_n)$ by the hedging property of ambiguity rules while $\rho(r_i) = \rho^1(r_i)$ for $1 \leq i \leq n-1$. By strict concavity of the sender's utility function and since $(\sigma,\rho)$ is an equilibrium, we necessarily have $m' \in M^1_n$ with a strict incentive to send $r_{n-1}$ instead of $r_n$, inducing $\rho^1(r_{n-1})$, which fulfills $\rho^1(r_{n-1}) = \rho(r_{n-1}) < \rho(r_n) \leq \rho^1(r_n)$. %
We may assume that $m'$ is minimal in $M^1_n$ w.r.t.\ the order of models: if there is some $m'' \in M^1_n$  without a profitable deviation, i.e., $U_{m'',h}(\rho^1(r_n),b) \geq U_{m'',h}(\rho(r_{n-1}),b)$, then $U_{m''',h}(\rho^1(r_n),b) > U_{m''',h}(\rho^1(r_{n-1}),b)$ by strict single crossing for all subsequent models $m''<m'''$ in $M^1_n$. Denote by $\m'$ the set of all minimal elements of $M^1_n$ with this profitable deviation.\footnote{Indeed, we could even choose any non-empty subset of $\m'$ in the following.}%

In the next step, we consider the modified strategy $\sigma^2$ defined by the partition $M^2_{n-1} = M^1_{n-1} \cup \m'$, $M^2_{n} = M^1_n \setminus \m'$ and $M^2_i = M^1_i$ for all $i \neq n-1,n$. Use again the reports $r_1, \ldots, r_n$ accordingly. %
Let $\rho^2$ be the receiver's unique best reply to $\sigma^2$, which by the hedging property of ambiguity rules fulfills $\rho^1(r_n) \leq \rho^2(r_n)$, $\rho^1(r_{n-1}) \leq \rho^2(r_{n-1})$, and $\rho^1(r_i) = \rho^2(r_i)$ for all $i \neq n-1,n$. In particular $\rho^1(r_i) \leq \rho^2(r_i)$ for all $1 \leq i \leq n$. %
If $(\sigma^2,\rho^2)$ is still no equilibrium, we iteratively change the sender strategy according to a profitable deviation as before.

We claim that this process eventually terminates at an $n$-step equilibrium. To this end, suppose $\sigma^{j-1}$ induces $n$ distinct optimal receiver actions and has no profitable deviation inducing a higher report for any model. We establish the following two facts for a modification $\sigma^j$ resulting from profitable deviations of $\sigma^{j-1}$ by adding a set $\m'$ of minimal elements of some $M_{i^{j-1}}$, $2 \leq i^{j-1} \leq n$, to $M_{i^{j-1}-1}$ if $(\sigma^j,\rho^j)$ is no equilibrium:
\begin{enumerate}[(i)]
    \item\label{item: thm: shift right} There is no profitable deviation from $\sigma^j$ by sending a higher report while there is a profitable deviation sending a lower report for some model that is minimal in its partition element: %
We have $\rho^{j-1}(r_i)\leq \rho^{j}(r_i)$ for all $1 \leq i \leq n$ since $\sigma^j$ arose from adding a set $\m'$ of minimal elements of $M^{j-1}_{i^{j-1}}$ to $M^{j-1}_{i^{j-1}-1}$. Assume by means of contradiction that there is a profitable deviation from $\sigma^j$ sending the next higher report $r_{i+1}$ for some model $m'' \in M^j_i$. Note that $m''$ cannot be an element of $\m'$ since the receiver actions have weakly increased from $\rho^{j-1}$ to $\rho^j$, hence $m'' \in M^{j-1}_i$. We proceed by cases: If $\rho^{j-1}(r_{i+1}) \leq a(m'',h,b)$, there must have been such a profitable deviation for $\sigma^{j-1}$ since $\rho^{j-1}(r_i) < \rho^{j-1}(r_{i+1})\leq a(m'',h,b)$. If $a(m'',h,b)< \rho^{j-1}(r_{i+1})$ and under $\sigma^j$ sending $r_{i+1}$ (inducing $\rho^j(r_{i+1}) \geq \rho^{j-1}(r_{i+1})$) is strictly better than sending $r_i$ (inducing $\rho^{j}(r_i) \geq \rho^{j-1}(r_i)$) this must also be the case under $\sigma^{j-1}$ which is excluded by the assumptions on $\sigma^{j-1}$. Consequently, a profitable deviation can only occur by sending a lower report for some model $m'' \in M^j_{i^j}$, $2 \leq i^j \leq n$. As explained above, we may assume that this is minimal in $M^j_{i^j}$ and thus yield again a consecutive partition.
\item\label{item: thm: different actions} $\rho^j$ gives rise to $n$ distinct receiver actions: %
Assume by means of contradiction that $\rho^{j-1}(r_i) < \rho^{j-1}(r_{i+1})$, but $\rho^{j}(r_i) = \rho^{j}(r_{i+1}) =: a'$ at some iteration step $j$ for some $i$. By construction, $M^j_i = M^{j-1}_i \cup \m'$, $M^j_{i+1} = M^{j-1}_{i+1} \setminus \m'$ for a set $\m' \subseteq M^{j-1}_{i^{j-1}}$ of models with profitable deviations. By the hedging property of an ambiguity rule, we must have $a(m',h,0) = a' = a(m^{j}_{i+1},h,0)$ for any minimal $m^{j}_{i+1} \in M^{j}_{i+1}$ and any $m' \in \m'$. Again by the hedging property, we find $\rho^{j-1}(r_{i+1}) = a(M^{j-1}_{i+1},h) = a'$ as $M^{j-1}_{i+1} = M^{j}_{i+1} \cup \m'$. Consider any $m' \in \m' \subseteq M^{j-1}_{i+1}$ and note that $\rho^{j-1}(r_i)<\rho^{j-1}(r_{i+1}) = a' = a(m',h,0)< a(m',h,b)$, contradicting that the iteration step $j-1 \mapsto j$ was a profitable deviation by strict concavity of $U_{m',h}(\cdot,b)$.
\end{enumerate}

Finally, we show that the algorithm terminates: %
Suppose $(\sigma^j,\rho^j)$ keeps not producing an equilibrium. By \eqref{item: thm: shift right}, the deviations implemented in the algorithm each involve assigning a set of models $\m'$ from some partition element $M^{j-1}_i$ to $M^{j-1}_{i-1}$ without depleting a partition due to \eqref{item: thm: different actions}. If this process goes on, $M^j_n$---and, iterating the following argument, in fact each of $M^j_2, \ldots, M^j_n$---will eventually consist only of models inducing the same bliss point, i.e., $\# \{ a(m_i, h, 0) \mid m_i \in M^j_i\} = 1$: The reason for this is that $M$ is finite and as soon as all models of $M_n^j$ have the same bliss point, say, $a(m_n,h,0)$, we have $\rho(r_n) = a(m_n,h,0)<a(m_n,h,b)$ so that only upward deviations can be profitable. But these do not exist due to \eqref{item: thm: shift right}. If the algorithm does not produce an equilibrium along the way, it will thus eventually reach the above configuration and there cannot be any further profitable deviation, thus constituting an equilibrium.\qedhere
\end{proof}

\begin{proof}[Proof of Corollary \ref{Corollary:comp_stat}]
    Applying arguments from the algorithm, we show how an $n$-step equilibrium $(\sigma,\rho)$ under $b$ can be transformed into an $n$-step equilibrium under $b'$: Start with $\sigma$ and the receiver's best reply $\rho'$ under $b'$. If $(\sigma, \rho')$ is also an equilibrium under $b'$, we are done. If not, the sender has an incentive to deviate. As $b'<b$, a profitable deviation entails sending a message $r_{i}$ that induces a smaller action $\rho'(r_i)$ for a left-most model $m' \in M_{i+1}$. If the resulting sender strategy together with the receiver's best reply to it does not constitute an equilibrium, we argue similarly to the proof of Theorem \ref{Theorem: N step equilibrium} that any profitable deviation is of the type described above, that the step-size does not go down and that the algorithm terminates at an equilibrium.\qedhere
\end{proof}

\begin{proof}[Proof of Proposition \ref{pro:fully_inf}]
\begin{enumerate}[(i)]
\item Consider a fully informative equilibrium $(\sigma,\rho)$, i.e., $\rho(\sigma(m)) = a(m,h,0)$ for all $m \in M$. %
        For any $m^T \in M$ let $a^T := a(m^T,h,0)$ and $a^+ := \min \{ a = a(m,h,0) \mid m \in M, a^T < a \}$ and define $W_{m^T,h} \colon \R_{\geq 0} \to \R, b \mapsto U_{m^T,h}(a^T,b)-U_{m^T,h}(a^+,b)$, which is a continuous function. Note that the sender has an incentive to deviate from revealing $m^T$ if and only if $W_{m^T,h}(b) < 0$ as $\rho(\sigma(m^T)) = a(m^T,h,0) < a(m^T,h,b)$ by supermodularity. Define
        \begin{equation*}
            \underline{b}(m^T,h) := \min \left(\{ b \geq 0 \mid W_{m^T,h}(b) = 0 \} \cup \{\infty\}\right).\label{eq: b lower bar mT h}
        \end{equation*}
        Since $W_{m^T,h}(0)>0$ we have $\underline{b}(m^T,h) > 0$. Now set $\underline{b}(h) := \min\{ \underline{b}(m^T,h) \mid m^T \in M \colon a(m^T,h,0) < \max_{m \in M} a(m,h,0)\}$. We have $\underline{b}(h)>0$ and for all $b \leq \underline{b}(h)$ the sender has no profitable deviation from the fully revealing strategy.        

\item $V(m^T,h,b)$ is increasing in $b$ by supermodularity. Under a large conflict of interest, we let $\overline{b}(h)$ be the infimum of all $\hat{b}(h)$ with $V(m^T,h,b)>0$ for all $b > \overline{b}(h)$ and all non-maximal $m^T$. For $b > \overline{b}(h)$, any equilibrium is babbling while for $b \leq \overline{b}(h)$ there exists a $2$-step equilibrium.\qedhere
\end{enumerate}
\end{proof}

\begin{proof}[Proof of Proposition \ref{Proposition: MLEU equilibria prefer informativeness}]
    Fix any $h$ and let $m^T$ be arbitrary and $\succ_{M,h}$ be the strict ordering on $M$ of the considered MLEU rule. In order to show that the sender is better off in $(\sigma,\rho)$, it suffices to show that the sender can induce the action $\rho'(\sigma'(m^T))$ in equilibrium $(\sigma,\rho)$ as well. To this end, assume first that $\sigma$ is more Blackwell informative than $\sigma'$ (cf.\ Remark \ref{Remark: more informative from equilibrium to Blackwell}). Define $M^{\sigma'}_{m^T} := (\sigma')^{-1}(\sigma'(m^T)) = \{ m \in M \mid \sigma'(m) = \sigma'(m^T) \}$ and let $m^*$ denote its maximal element w.r.t.\ $\succ_{M,h}$. In particular, $\rho'(\sigma'(m^T)) = a(M^{\sigma'}_{m^T},h) = a(m^*,h,0)$. Since $\sigma$ is more Blackwell informative than $\sigma'$, we have $M^{\sigma}_{m^*} := \sigma^{-1}(\sigma(m^*)) \subseteq M^{\sigma'}_{m^T}$ and thus $\rho(\sigma(m^*)) = a(M^{\sigma}_{m^*},h) = a(m^*,h,0)$, as desired. Assume now that $(\sigma,\rho)$ is more informative than $(\sigma',\rho'
    )$. Since the statement only concerns the action-model profiles, we can without loss assume that both equilibria are reduced. In that case, $\sigma$ is more Blackwell informative than $\sigma'$ by Remark \ref{Remark: more informative from equilibrium to Blackwell}, letting us apply the above argument to conclude the proof.\qedhere
\end{proof}

\begin{proof}[Proof of Proposition \ref{pro:extreme_histories}]
    Consider any $h^\Sigma$. By the Reduction Lemma \ref{Lemma: reduction lemma}, we can identify the set of models with the set of induced actions (and beliefs) under $h^\Sigma$,
    \begin{equation*}
        \mathcal{A}(h^\Sigma)=\left\{a(m,h',0) \mid m\in M, \sum_{k=1}^K h_k'=h^\Sigma\right\}=\left\{\frac{\sum_{i\in m}h_i'+1}{\#m+2} \mid m\in M, \sum_{k=1}^K h_k'=h^\Sigma\right\}.
    \end{equation*}  
    Note that $\underline b(h^\Sigma)=\frac{1}{2}%\min_{a,a' \in \mathcal{A}(h^\Sigma): a \neq a'} |a-a'| 
    \min \{ |a-a'| \mid {a,a' \in \mathcal{A}(h^\Sigma): a \neq a'}\}$, and thus $\underline{b}(h^\Sigma)$ is independent of the ambiguity rule. Symmetry then follows since, for any given $k\in\{1,2,\ldots,K\}$,
        \begin{align*}
          \frac{s+1}{k+2}\in \mathcal{A}(h^\Sigma)\iff& \max\{0,h^\Sigma-K+k\}\leq s\leq \min\{k,h^\Sigma\}\\
        \iff& \max\{0,K-h^\Sigma-K+k\}\leq k-s\leq \min\{k,K-h^\Sigma\}\\
        \iff& \frac{k-s+1}{k+2}\in \mathcal{A}(K-h^\Sigma)%\\
        \iff%&
        1-\frac{s+1}{k+2}\in \mathcal{A}(K-h^\Sigma).
        \end{align*}
    For the second claim, note first that the smallest common denominator of distinct $a,a'\in\mathcal{A}(h^\Sigma)$ is at most $(K+1)(K+2)$, which implies $ |a-a'|\geq \frac{1}{(K+1)(K+2)}$. %
    This inequality is an equality only if $a=\frac{s+1}{K+2}$ and $a'=\frac{s'+1}{K+1}$ (or vice versa) for $0\leq s\leq K$ and $0\leq s'\leq K-1$. Next, observe that
    \begin{align*}
      |a-a'|
      =\frac{1}{(K+1)(K+2)}\iff&\left|(s+1)(K+1)-(s'+1)(K+2)\right|=1\\
    \iff&\left|s(K+1)-s'(K+2)-1\right|=1\\
    \iff&s'(K+2)=s(K+1) \vee s=s'+\frac{2+s'}{K+1}\\
    \iff&s=s'=0 \vee (s=K \wedge s'=K-1),
        \end{align*}
        where the last step follows from $s(K+1)>(s-1)(K+2)$ for all $s\leq K$ and $$\left(\frac{2+s'}{K+1}\in\N \wedge 0\leq s'\leq K-1\right) \iff s'=K-1.$$
        The claim follows since $\frac{1}{K+2}\in \mathcal{A}(h^\Sigma)\iff h^\Sigma=0$ and $\frac{K+1}{K+2}\in \mathcal{A}(h^\Sigma)\iff h^\Sigma=K$.
    \end{proof}

\begin{proof}[Proof of Proposition \ref{prop:partial_asym}]
Fix any $K \geq 2$ and consider first the MLEU ambiguity rule. Consider $h^\Sigma=K$, i.e., $h = \vec{1}$, and note that in this case $a(m,h,0)=\frac{\#m+1}{\#m+2}$ for any $m$. Thus, the expected fit of model $m$ is
\begin{align*}
\Pr(h \mid m,F_0)=\int_0^1\theta^{\#m}\left(\frac{1}{2}\right)^{K-\#m}d\theta=\left(\frac{1}{2}\right)^{K-\#m}\frac{1}{\#m+1},
\end{align*}
which is (strictly) increasing in $\#m$ (if $\#m\geq 1$).
Moreover, any $2$-step partition can be written as
$M^{\vec{1}}_1(k) := \{m \mid \#m\leq k\}$ and $M^{\vec{1}}_2(k) = \{m \mid \#m> k\}$ for some $k\in\{0,1,\ldots,K-1\}$; the induced actions then are $a(M^{\vec{1}}_1(k),\vec{1})=\frac{k+1}{k+2}$ and $a(M^{\vec{1}}_2(k),\vec{1})=\frac{K+1}{K+2}$. %
Observe that since $a(m,h,b)>a(m,h,0)$, it is sufficient to consider upward deviations, i.e., whether the sender would like to induce $a(M^{\vec{1}}_2(k),\vec{1})$ rather than $a(M^{\vec{1}}_1(k),\vec{1})$ when $m^T\in M^{\vec{1}}_1(k)$.  The sender has no incentives to deviate upward if and only if she has no incentives to deviate when $m^T\in\{m| \#m=k\}$ (then $a(m,h,b)=\frac{k+1}{k+2}+b$), which is equivalent to 
\begin{align*}
a(m,h,b)-a(M^{\vec{1}}_1(k),\vec{1})\leq a(M^{\vec{1}}_2(k),\vec{1})-a(m,h,b)
\iff& b\leq\frac{1}{2}\left(\frac{K+1}{K+2}-\frac{k+1}{k+2}\right).
\end{align*}
Thus, we obtain $\bar b(h)=\max_{k=0,1,\ldots,K-1} \frac{1}{2}\left(\frac{K+1}{K+2}-\frac{k+1}{k+2}\right)=\frac{K}{4(K+2)}$.

Next, consider $h^\Sigma=0$, i.e., $h = \vec{0}$. Analogously to the first case, $a(m,h,0)=\frac{1}{\#m+2}$ for any $m$ and the expected fit of model $m$ is (strictly) increasing in $\#m$ (if $\#m\geq 1$). Therefore, any $2$-step partition can be written as $M^{\vec{0}}_1(k) := \{m \mid \#m\geq k\}$ and $M^{\vec{0}}_2(k) = \{m \mid \#m< k\}$ for some $k\in\{1,2,\ldots,K\}$; the induced actions then are $a(M^{\vec{0}}_1(k),\vec{0})=\frac{1}{K+2}$ and $a(M^{\vec{0}}_2(k),\vec{0})=\frac{1}{k+1}$. The sender has no incentives to deviate upward if and only if she has no incentives to deviate when $m^T\in\{m| \#m=k\}$, which is equivalent to 
\begin{align*}
&a(m,h,b)-a(M^{\vec{0}}_1(k),\vec{0})\leq a(M^{\vec{0}}_2(k),\vec{0})-a(m,h,b)\\
\iff&b\leq \frac{1}{2(k+1)}+\frac{1}{2(K+2)}-\frac{1}{k+2}.
\end{align*}
Thus, we obtain $\bar b(h)=\max_{k=1,2,\ldots,K} \frac{1}{2(k+1)}+\frac{1}{2(K+2)}-\frac{1}{k+2} = \frac{1}{2(K+1)(K+2)}$. %
Finally, a fully informative strategy is part of an equilibrium if and only if  
\begin{align*}
b\leq \frac{1}{2}\min_{m,m'\in M:a(m,h,0)\neq a(m',h,0)} |a(m,h,0) -a(m',h,0)|
=\frac{1}{2(K+1)(K+2)}.
\end{align*}

Second, consider the MEU ambiguity rule. Consider $h^{\Sigma} = 0$, i.e., $h = \vec{0}$, and recall that in this case $a(m, h,0) = \tfrac{1}{\#m+2}$ for any $m$. By considering $k \in \{1,2 \ldots, K\}$, we go through all bipartitions in the order of their bliss points by setting $M^{\vec{0}}_1(k) := \{m \mid \#m\geq k\}$ and $M^{\vec{0}}_2(k) = \{m \mid \#m< k\}$. We now state the optimal actions under the MEU ambiguity rule for the relevant partition elements, providing a derivation at the end of this proof.
\begin{align}
    a(M^{\vec{0}}_1(k),\vec{0}) &= \min\left\{ \frac{1}{k + 2}, \frac{K+k+5}{(K+3)(k+3)} \right\},\label{eq: optimal action MEU M1}\\
        a(M^{\vec{0}}_2(k), \vec{0})&= 
        \min\left\{ \frac{1}{2}, \frac{k + 4}{3(k + 2)} \right\}.\label{eq: optimal action MEU M2}
\end{align}
The sender has no incentives to deviate upward if and only if she has no incentives to deviate when $m^T\in\{m| \#m=k\}$ (then $a(m,h,b)=\frac{1}{k+2}+b$), which is equivalent to $\betrag{\tfrac{1}{k+2}+ b - a(M^{\vec{0}}_1(k),\vec{0}) } \leq \betrag{\tfrac{1}{k+2}+ b - a(M^{\vec{0}}_2(k),\vec{0})}$. Thus, with \eqref{eq: optimal action MEU M1} and \eqref{eq: optimal action MEU M2} we obtain
\begin{align*}
   \overline{b}(0) &= \max_{k=1,2,\ldots,K} \frac{a(M^{\vec{0}}_1(k),\vec{0}) + a(M^{\vec{0}}_2(k),\vec{0})}{2} - \frac{1}{k+2}\\
   &= \max_{k=1,2,\ldots,K} \frac{1}{2}\left(\min\left\{ \frac{1}{k + 2}, \frac{K+k+5}{(K+3)(k+3)} \right\} + \min\left\{\frac{1}{2}, \frac{k + 4}{3(k + 2)}\right\}\right) - \frac{1}{k+2}\\
   & =\frac{K+1}{6(K+2)}.
\end{align*}

We now turn towards $h^{\Sigma} = K$, i.e., $h = \vec{1}$. Recall that in this case $a(m,h,0) = \tfrac{\#m+1}{\#m+2}$. The subsequent analysis can be based on the previous calculations by observing the following symmetry to the case $h= \vec{0}$: define the bipartition via $M^{\vec{1}}_1(k) := \{m \mid \#m< k\}$ and $M^{\vec{1}}_2(k) = \{m \mid \#m\geq k\}$ for $k \in \{1,2 \ldots, K\}$. The expected utility functions for the singleton models are
\begin{align*}
    U_{\{m\}, \vec{0}}(a) &= - (a - \tfrac{1}{\#m+2})^2 - \frac{\#m+1}{(\#m+2)^2(\#m+3)}\\
    U_{\{m\}, \vec{1}}(a) &= - (a - \tfrac{\#m+1}{\#m+2})^2 - \frac{\#m+1}{(\#m+2)^2(\#m+3)},\nonumber
\end{align*}
i.e., parabolas the maxima of which have the relation $a(\{m\},\vec{0}) = 1-a(\{m\},\vec{1})$ and who share the same height in their maxima. Consequently, also $a(\tilde{M}, \vec{0}) = 1-a(\tilde{M}, \vec{1})$ for each $\tilde{M} \subseteq \{0, \ldots, K\}$. In analogy to before, we thus see that
\begin{align*}
    \overline{b}(K)&= \max_{k=1,2,\ldots,K}\frac{a(M^{\vec{1}}_1(k),\vec{1}) + a(M^{\vec{1}}_2(k),\vec{1})}{2} - \frac{k}{k+1}\\
    %&= \frac{1- a(M^{\vec{0}}_1(k),\vec{0}) + 1 - a(M^{\vec{0}}_2(k),\vec{0})}{2} - \frac{k}{k+1}\\
    &= \max_{k=1,2,\ldots,K}\frac{1}{k+1} - \frac{a(M^{\vec{0}}_1(k),\vec{0}) +  a(M^{\vec{0}}_2(k),\vec{0})}{2}\\
     &= \max_{k=1,2,\ldots,K}\frac{1}{k+1} - \frac{1}{2}\left(\min\left\{ \frac{1}{k + 2}, \frac{K+k+5}{(K+3)(k+3)} \right\} + \min\left\{\frac{1}{2}, \frac{k + 4}{3(k + 2)}\right\}\right)\\
     &= \max\left\{\frac{1}{12},\frac{K}{8(K+3)} \right\}.
\end{align*}

Finally, it is left to derive the optimal actions under the MEU ambiguity rule \eqref{eq: optimal action MEU M1} and \eqref{eq: optimal action MEU M2}. Consider the following fact. Let $f_i(x):=f_{\alpha_i,\beta_i}(x) = -(x-\alpha_i)^2 - \beta_i$ for real numbers $\alpha_1 > \ldots > \alpha_n$ and $\beta_1 \geq \ldots \geq \beta_n$. Note that these functions are all strictly quasi-concave: $f_{i}$ is strictly increasing for $x <\alpha_i$ and strictly decreasing for $\alpha_i < x$. The minimum function $f := \min\{ f_1, \ldots, f_n\}$ is also strictly quasi-concave with unique maximizer $\alpha^* := \argmax_x f(x)$. %
For $x> \alpha_1$, $f$ is strictly decreasing as all $f_i$ are, thus $\alpha^* \leq \alpha_1$. Likewise, $\alpha_n \leq \alpha^*$. %
Note that $\alpha^*$ is either one of the $\alpha_i$ or marks the intersection of at least two functions. By assumption on the $\beta_i$'s and the previous insight, $\alpha^*$ is thus either $\alpha_1$ or an intersection on the interval $[\alpha_n, \alpha_1]$. %
If $\alpha^*$ is an intersection of a set $I \subseteq \{1,\ldots,n\}$ of functions, there must be $i,j \in I$ with $\alpha_i \leq \alpha^* \leq  \alpha_j$, since otherwise $f$ was either strictly decreasing or increasing in an open neighborhood of $\alpha^*$ (the set of intersections is finite and thus contains no accumulation points). %
Now, observe that for each $i \in \{2, \ldots, n\}$ and $x\leq \alpha_i$ we have $f_1(x) < f_i(x)$, for the functions in question, since $\alpha_i < \alpha_1$, $f_1(\alpha_1) = -\beta_1 \leq -\beta_i =  f_i(\alpha_i)$ and $\tfrac{\partial}{\partial x} f_1 = -2x + 2 \alpha_1 > -2x + 2 \alpha_i = \tfrac{\partial}{\partial x} f_i$. %
Together, the last two insights reveal that if $\alpha^*$ is an intersection point, it must intersect $f_1$. As a result, $\alpha^*$ is the minimum among $\alpha_1$ and the intersection points of $f_1$ with the other $f_i$. %
In our concrete case, $f_k(x) := - (x - \tfrac{1}{k+2})^2 - \tfrac{k+1}{(k+2)^2(k+3)}$. If we consider $m' \in M^{\vec{0}}_1(k) = \{m \mid \#m\geq k\}$ and let $k'=\#m'$, the intersection of $f_{k'}$ with $f_{k}$ is $\tfrac{k'+k+5}{(k+3)(k'+3)}$, which is strictly decreasing in $k'$ and thus minimized for $k' = K$. With $\alpha_1 = \tfrac{1}{k + 2}$ we hence obtain \eqref{eq: optimal action MEU M1}. %
Similarly for $m' \in M^{\vec{0}}_2(k) = \{m \mid \#m< k\}$ with $k'=\#m'$ we find the intersections at $\tfrac{k'+5}{3(k'+3)}$ which is minimized by $k' = k-1$. With $\alpha_1 = \tfrac{1}{2}$ we thus obtain \eqref{eq: optimal action MEU M2}.\qedhere
\end{proof}

\begin{proof}[Proof of Lemma \ref{Lemma: naive easier to convince}]
        Let $m^T \in \Meq$. Then $\rho(\sigma(m^T)) \neq a(m^T,h,0)$. Since we consider an MLEU equilibrium, we have $\rho(\sigma(m^T)) = a(m^*,h,0)$ for some $m^* \in \sigma^{-1}(\sigma(m^T))$ with $\Pr(h \mid m^*,F_0) \geq \Pr(h \mid m^T,F_0)$. Thus, $m^T \in \Mna$.
\end{proof}

\begin{proof}[Proof of Proposition \ref{Proposoition: naive easier to convince}]
    By inequality \eqref{persuasive_power}, $(\sigma^\dagger,\rho^\dagger)$ has more persuasive power than $(\sigma,\rho)$. If now $m^T \in \Mna \setminus \Meq$, we have strictly more persuasive power as
    $$U_{m^T,h}(\rho(\sigma(m^T)),b) \leq U_{m^T,h}(a(m^T,h,0),b) < U_{m^T,h}(\rho^\dagger(\sigma^\dagger(m^T)),b).$$
\vskip-8mm
\end{proof}
    
\begin{proof}[Proof of Proposition \ref{Proposition: generic indifference, improve upon naive}]
    First, let $b \leq \underline{b}(h)$. At any true model $m^T$, the best receiver bliss point the sender can induce is then $a(m^T,h,0)$. Consequently, $\Meq = \Mna = \emptyset$. %
    Second, if $b > \overline{b}(h)$, provided preferences allow for large conflicts of interest, the only equilibrium is babbling, inducing the constant $a(M,h)$. For every true model $m^T$ with $a(M,h) \leq a(m^T,h,0)$, the sender's utility thus stays at $U_{m^T,h}(a(M,h),b) \leq U_{m^T,h}(a(m^T,h,0),b)$ in equilibrium. If there exists $m^*$ with $m^T < m^*$ and $\Pr(h \mid m^*,F_0) \geq \Pr(h \mid m^T,F_0)$, a na\"ive receiver can be convinced to take action $a(m^*,h,0)$ under the true model $m^T$, which strictly improves the sender's expected utility.\qedhere
\end{proof}
\end{document}